\documentclass[twoside,twocolumn,9pt,a4]{article}
\usepackage{extsizes}
\usepackage[super,sort&compress,comma]{natbib} 
\usepackage[version=3]{mhchem}
\usepackage[left=1.5cm, right=1.5cm, top=1.785cm, bottom=2.0cm]{geometry}
\usepackage{balance}
\usepackage{mathptmx}
\usepackage{sectsty}
\usepackage{graphicx} 
\usepackage{lastpage}
\usepackage[format=plain,justification=justified,singlelinecheck=false,font={stretch=1.125,small,sf},labelfont=bf,labelsep=space]{caption}
\usepackage{float}
\usepackage{fancyhdr}
\usepackage{fnpos}
\usepackage[english]{babel}
\addto{\captionsenglish}{%
  
}
\usepackage{array}
\usepackage{droidsans}
\usepackage{charter}
\usepackage[T1]{fontenc}
\usepackage[usenames,dvipsnames]{xcolor}
\usepackage{setspace}
\usepackage[compact]{titlesec}
\usepackage{hyperref}

\newcommand{\rev}[1]{\textcolor{black}{#1}}

\usepackage{epstopdf}

\definecolor{cream}{RGB}{222,217,201}

\usepackage{multirow,booktabs,subcaption,amsmath,amsfonts,amssymb,bm,tikz,tikz-3dplot}
\begin{document}

\pagestyle{fancy}
\thispagestyle{plain}
\fancypagestyle{plain}{
\renewcommand{\headrulewidth}{0pt}
}

\makeFNbottom
\makeatletter
\renewcommand\LARGE{\@setfontsize\LARGE{15pt}{17}}
\renewcommand\Large{\@setfontsize\Large{12pt}{14}}
\renewcommand\large{\@setfontsize\large{10pt}{12}}
\renewcommand\footnotesize{\@setfontsize\footnotesize{7pt}{10}}
\makeatother

\renewcommand{\thefootnote}{\fnsymbol{footnote}}
\renewcommand\footnoterule{\vspace*{1pt}%
\color{cream}\hrule width 3.5in height 0.4pt \color{black}\vspace*{5pt}} 
\setcounter{secnumdepth}{5}

\makeatletter 
\renewcommand\@biblabel[1]{#1}            
\renewcommand\@makefntext[1]%
{\noindent\makebox[0pt][r]{\@thefnmark\,}#1}
\makeatother 
\renewcommand{\figurename}{\small{Fig.}~}
\sectionfont{\sffamily\Large}
\subsectionfont{\normalsize}
\subsubsectionfont{\bf}
\setstretch{1.125} 
\setlength{\skip\footins}{0.8cm}
\setlength{\footnotesep}{0.25cm}
\setlength{\jot}{10pt}
\titlespacing*{\section}{0pt}{4pt}{4pt}
\titlespacing*{\subsection}{0pt}{15pt}{1pt}

\fancyfoot{}
\fancyfoot[RO]{\footnotesize{\sffamily{1--\pageref{LastPage} ~\textbar  \hspace{2pt}\thepage}}}
\fancyfoot[LE]{\footnotesize{\sffamily{\thepage~\textbar\hspace{3.45cm} 1--\pageref{LastPage}}}}
\fancyhead{}
\renewcommand{\headrulewidth}{0pt} 
\renewcommand{\footrulewidth}{0pt}
\setlength{\arrayrulewidth}{1pt}
\setlength{\columnsep}{6.5mm}
\setlength\bibsep{1pt}

\makeatletter 
\newlength{\figrulesep} 
\setlength{\figrulesep}{0.5\textfloatsep} 

\newcommand{\topfigrule}{\vspace*{-1pt}%
\noindent{\color{cream}\rule[-\figrulesep]{\columnwidth}{1.5pt}} }

\newcommand{\botfigrule}{\vspace*{-2pt}%
\noindent{\color{cream}\rule[\figrulesep]{\columnwidth}{1.5pt}} }

\newcommand{\dblfigrule}{\vspace*{-1pt}%
\noindent{\color{cream}\rule[-\figrulesep]{\textwidth}{1.5pt}} }

\makeatother

\twocolumn[
  \begin{@twocolumnfalse}
\vspace{1em}
\sffamily

\noindent\LARGE{\textbf{Controllable particle migration in liquid crystal flows$^\dag$}} \\

\noindent\large{Magdalena Lesniewska\textit{$^{a}$}, Nigel Mottram\textit{$^{b}$}, and Oliver Henrich\textit{$^{a \ddag}$}} \\

\noindent\normalsize{We observe novel positional control of a colloidal particle in microchannel flow of a nematic liquid crystal. Lattice Boltzmann simulations show multiple equilibrium particle positions, the existence and position of which are tunable using the driving pressure, in direct contrast to the classical Segr\'e-Silberberg effect in isotropic liquids. In addition, particle migration in nematic flow occurs an order of magnitude faster. These new equilibria are determined through a balance of elastic forces, hydrodynamic lift and drag as well as order-flow interactions through the defect structure around the particle.}\\


\end{@twocolumnfalse} \vspace{0.6cm}

  ]

\renewcommand*\rmdefault{bch}\normalfont\upshape
\rmfamily
\section*{}
\vspace{-1cm}


\footnotetext{\textit{$^{a}$~Department of Physics, University of Strathclyde, Glasgow G4 0NG, UK.}}
\footnotetext{\textit{$^{b}$~School of Mathematics \& Statistics, University of Glasgow, Glasgow G12 8QQ, UK. }}
\footnotetext{\ddag~Email: oliver.henrich@strath.ac.uk}





\section{Introduction}

Particle-laden flows are at the heart of many scientific and engineering applications, ranging from the dispersion of pollutants and rain formation in the atmosphere to pharmaceutical aerosols and industrial spray applications to geological sedimentation and combustion processes, and even planetary formation. Such flows consist of a continuous liquid host phase and dispersed particles. 

At low particle concentrations these systems are generally modelled by the Navier-Stokes equations, with suitable modifications to account for the momentum exchange between the host liquid and dispersed particles \cite{Batchelor}. In spite of their relative simplicity, particle-laden flows show non-trivial behaviour, such as the preferential migration of the particles to specific regions. Segr\'e and Silberberg \cite{Segre1961,Segre1962} were among the first to report this phenomenon. In capillary flow of isotropic fluids, they observed that neutrally buoyant dispersed particles migrate to a single radial position approximately $0.6$ radii from the capillary axis, while being advected downstream. \rev{This effect is generally attributed to a balance between a shear-induced inertial lift force that drives the particles towards the walls and regions of higher velocity gradient, and a pressure-induced force that increases as the particles move closer to the walls and acts in the opposite direction.} A qualitative explanation \cite{Saffman1965,Gotoh1970}, a subsequent quantitative analysis \cite{Ho1974, Schonberg1989} and then an experimental extension to higher Reynolds numbers \cite{Matas2004} and different tube geometries \cite{DiCarlo2007} have already been provided.

Much less is known about the dynamics and migration of particles in liquid crystalline host phases, whose internal structure offers additional possibilities in terms of elastic and order-driven control. Theoretical studies have considered the Stokes drag of a particle in a nematic host \cite{Stark2002} or particle aggregates \cite{Mondal2018, Mondal2018a}, and simulation studies have investigated flow-order tensor interactions around colloidal particles \cite{Stieger2014} for fixed or simply advected particles. 
Experimental studies of nematics in microfluidic geometries have been concerned with pure phases \cite{Sengupta2013, Sengupta2013b, Sengupta2014, Anderson2015, Giomi2017, Copar2020, Steffen2021, Copar2021}.
However, recent experimental research has also considered particle-laden nematic flows which contain defects and/or obstacles \cite{Chen2017}, or include electro-hydrodynamic \cite{Sasaki2015} or electro-osmotic flow effects \cite{Lavrentovich2010, Lavrentovich2014, Peng2018}.
\rev{Further studies investigated the microfluidic flow of liquid crystals in more exotic settings, for instance the sedimentation of discoidal particles in cholesteric finger textures \cite{Chen2015} or modified free energy landscapes with alternating splay and bend distortions that were introduced through wavy walls \cite{Luo2018}. More recently, active nematics were studied in a network of connected annular microfluidic channels \cite{Hardouin2020}.}

Here, we report results of 3D simulations that demonstrate novel positional control of a colloidal particle in microchannel flow of a nematic liquid crystal, which serves as a prototype for anisotropic fluids with internal order structure. As well as multistability, we show that the equilibrium particle position in the dynamic system with respect to the channel centre or walls is tunable through the applied pressure gradient.

\section{Theory}\label{theory}

\subsection{Landau-de Gennes free energy}

The local order of the liquid crystal is described by a traceless and symmetric second-order tensor $\bm{Q}(\bm{r}, t)$. Its largest eigenvalue $q<2/3$ is referred to as the scalar order parameter and provides a measure of the liquid crystalline order at a certain position and time. The eigenvector, $\bm{d}$, associated with $q$ is called the director and describes the average orientation of the liquid crystal molecules at a certain position and time.

In equilibrium, the liquid crystal order is determined through minimisation of its free energy, commonly described by the Landau-de Gennes free energy functional 
\begin{equation}
{\cal F} = \int_V \rev{f}\, dV + \int_S \rev{f_s}\, dS,
\end{equation}
which includes the volume contribution $\rev{f=f_b + f_g}$, that itself consists of a bulk contribution $\rev{f_b}$ and a gradient contribution $\rev{f_g}$, and a surface contribution $\rev{f_s}$. The bulk free energy density is given by
\begin{eqnarray}
\rev{f_b}(\bm{Q})=
{\frac{A_0}{2}}\left(1 - \dfrac{\gamma}{3}\right) Q_{\alpha\beta}^2 -{\frac{A_0}{3}}\gamma \,Q_{\alpha\beta} Q_{\beta\pi} Q_{\pi\alpha}+ {\frac{A_0}{4}}  \gamma \,(Q_{\alpha\beta}^2)^2,
\label{eq-lc-fed-bulk}
\end{eqnarray}
where we use the Einstein summation convention, which implies that Greek indices that appear twice are summed over. $A_0$ is a constant that sets the overall energy scale and the parameter $\gamma$ controls the temperature difference from the isotropic-nematic transition, and is related to a reduced temperature $\tau$ by
\begin{equation}
\tau = \frac{27}{\gamma}\left(1-\frac{\gamma}{3}\right).
\end{equation}
For $\gamma>3$ the ordered, nematic state is the equilibrium phase, whereas for $2.7 \le \gamma \le 3$ the nematic state is metastable. For $\gamma<2.7$ the isotropic state is the equilibrium phase.
  
The gradient free energy density $f_g$ contains the contributions of splay, bend, splay and twist deformations of the director field,
\begin{eqnarray}
\rev{f_g}(\bm{Q}) =
{\frac{1}{2}} \kappa_0 (\partial_\alpha Q_{\alpha\beta})^2+ {\frac{1}{2}} \kappa_1 (\epsilon_{\alpha\mu\nu} \partial_\mu Q_{\nu\beta})^2,
\label{equation-lc-gradient-fe}
\end{eqnarray}
where $\partial_\alpha=\partial/\partial x_\alpha$ and $\epsilon_{\alpha\mu\nu}$ is the Levi-Civita symbol in three dimensions. In principle, the elastic constants $\kappa_0$, for splay and bend deformations, and $\kappa_1$, for twist deformations, can be different. However, in our simulations we use the one elastic constant approximation $\kappa_0=\kappa_1$.

The director is assumed to have a preferred normal orientation to the wall surfaces and to the surface of the colloidal particle, known as a homeotropic weak anchoring, and is described using a surface free energy term 
\begin{equation}
\rev{f_s}(\bm{Q}) = {\textstyle\frac{1}{2}} w (Q_{\alpha\beta} - Q_{\alpha\beta}^0)^2,
\label{surface_fe}
\end{equation}
where $w$ is the surface anchoring strength with values $w_{wall}$ and $w_{part}$ at the wall and particle surfaces, respectively, and the preferred orientation $Q^0_{\alpha\beta}$ is assumed uniaxial and is given by
\begin{equation}
Q^0_{\alpha\beta} = {\textstyle \frac{1}{2}} S_0 (3\,\rev{n_\alpha n_\beta} - \delta_{\alpha\beta}),
\end{equation}
where $\rev{\mathbf{n}}$ is the surface unit normal, $\delta_{\alpha\beta}$ is the Kronecker delta and $S_0$ is the preferred surface scalar order parameter given by
\begin{equation}
\label{equation-lc-amplitude}
S_0=\frac{2}{3}\left(\frac{1}{4}+\frac{3}{4}\sqrt{1 - \frac{8}{3\gamma}}\right).
\end{equation}

\subsection{Beris-Edwards model}

The time evolution of $Q_{\alpha\beta}$ is governed by the Beris-Edwards equation \cite{BerisEdwards} 
\begin{eqnarray}
\partial_t Q_{\alpha\beta} + \partial_\gamma (u_\gamma Q_{\alpha\beta})\qquad\nonumber&&\\
+ S_{\alpha\beta}(\bm{W},\bm{Q})& =& \Gamma\,  H_{\alpha\beta},
\label{equation-lc-beris-edwards}
\end{eqnarray}
where $\partial_t=\partial/\partial t$, $\bm{u}$ is the flow velocity, $\bm{H}$ is the molecular field, $\Gamma$ is a mobility parameter, 
and $\bm{S}(\bm{W},\bm{Q})$ denotes the response to shear and $\bm{W}$ is the velocity gradient tensor. The shear term is given by 
\begin{align}
S_{\alpha\beta}& (\bm{W},\,\bm{Q}) = \nonumber\\
& (\xi D_{\alpha\pi} + \Omega_{\alpha\pi})(Q_{\pi\beta} + {\textstyle \frac{1}{3}} \delta_{\pi\beta}) \nonumber\\
&\quad+ (Q_{\alpha\pi} + {\textstyle \frac{1}{3}} \delta_{\alpha\pi})(\xi D_{\pi\beta} - \Omega_{\pi\beta}) \nonumber\\
&\quad\quad- 2\xi(Q_{\alpha\beta} + {\textstyle\frac{1}{3}}\delta_{\alpha\beta})Q_{\pi\sigma}W_{\sigma\pi}
\end{align}
where $D_{\alpha\beta} = \frac{1}{2}(W_{\alpha\beta} + W_{\beta\alpha})$ and $\Omega_{\alpha\beta} = \frac{1}{2}(W_{\alpha\beta} - W_{\beta\alpha})$ are the symmetric and antisymmetric contributions to the velocity gradient tensor, respectively, and $\xi$ is the so-called flow-alignment parameter, a material constant representing an effective molecular aspect ratio which determines whether the liquid crystal molecules are in a flow-aligned state at the Leslie angle or tumbling state. 
The molecular field $\bm{H}$ is defined as a functional derivative of the free energy functional with respect to the order parameter,
\begin{equation}
H_{\alpha\beta} = -
  \frac{\delta \cal F} {\delta Q_{\alpha\beta}}
+ \frac{\delta_{\alpha\beta}}{3} {\rm Tr} \frac{\delta \cal F} {\delta Q_{\alpha\beta}}.
\label{molecular_field}
\end{equation}
The second term in Eq. \ref{molecular_field} involving the trace ensures tracelessness of the tensor order parameter as it evolves obeying Eq.\ref{equation-lc-beris-edwards}. \rev{This leads to the following molecular field:
\begin{eqnarray}
\label{eq-lc-h-full}
H_{\alpha\beta} = \hspace*{-0.6cm}&&-A_0 (1 - \gamma/3) Q_{\alpha\beta} + A_0 \gamma\, (Q_{\alpha\mu} Q_{\mu\beta} - {\textstyle\frac{1}{3}} Q_{\mu\nu}^2\delta_{\alpha\beta})\nonumber\\ 
\hspace*{-0.6cm}&&-A_0 \gamma\, Q_{\mu\nu}^2 Q_{\alpha\beta} +\kappa_0 \partial_\alpha \partial_\mu Q_{\mu\beta}
+ \kappa_1 \partial_\mu(\partial_\mu Q_{\alpha\beta} - \partial_\alpha Q_{\mu\beta})\nonumber\\
\end{eqnarray}}
The governing  equations of hydrodynamic motion are the equation of mass conservation, also known as the continuity equation, and the Navier-Stokes equation that describes the conservation of linear momentum. In tensor notation they read
\begin{equation}
\partial_t \rho + \partial_\alpha (\rho u_\alpha) = 0
\label{eq_mass1}
\end{equation}
and
\begin{align}
\hspace*{-0.2cm}\partial_t (\rho & u_\alpha) = \rev{\partial_\beta \Pi^{(LC)}_{\alpha\beta}+ \partial_\beta \Pi^{(HD)}_{\alpha\beta}},
\label{eq_momentum1}
\end{align}
respectively.
Eq. \ref{eq_mass1} relates the local rate of change of the density $\rho$ to the advection of mass by the fluid velocity $u_\alpha$. Eq. \ref{eq_momentum1} is Newton's second law of momentum change for the fluid and involves \rev{the thermotropic stress tensor $\Pi^{(LC)}_{\alpha\beta}$ and the hydrodynamic stress tensor $\Pi_{\alpha\beta}^{(HD)}$.}
The thermotropic stress arises due to the liquid crystal and is given by
\begin{equation}
\rev{\Pi^{(LC)}_{\alpha\beta}} = \sigma_{\alpha\beta} + \tau_{\alpha\beta} - \partial_\alpha Q_{\mu\nu} \frac{\delta {\cal F}}{ \delta \partial_\beta Q_{\mu\nu}}.
\label{equation-lc-stress}
\end{equation} 
In Eq.~\ref{equation-lc-stress}, $\sigma_{\alpha\beta}$ and $\tau_{\alpha\beta}$ are the symmetric and antisymmetric stress contributions, respectively, defined as
\begin{eqnarray}
\hspace*{-0.6cm}\sigma_{\alpha\beta} &=& \rev{-p_0}\, \delta_{\alpha\beta} - \xi H_{\alpha\mu}(Q_{\mu\beta} + {\textstyle\frac{1}{3}}\delta_{\mu\beta})- \xi (Q_{\alpha\mu} + {\textstyle\frac{1}{3}}\delta_{\alpha\mu}) H_{\mu\beta}\nonumber\\
&& + 2\xi (Q_{\alpha\beta} + {\textstyle \frac{1}{3}}\delta_{\alpha\beta}) Q_{\mu\nu} H_{\mu\nu},
\label{sym_stress}
\end{eqnarray}
where \rev{$p_0=-\left(\partial {\cal F} / \partial V\right)_T = -f$ is the isotropic pressure}, and
\begin{equation}
\tau_{\alpha\beta} = Q_{\alpha\mu} H_{\mu\beta} - H_{\alpha\mu} Q_{\mu\beta}.
\label{antisym_stress}
\end{equation}
The final term in   Eq. \ref{equation-lc-stress} is expanded as
\begin{equation}
\begin{split}
\partial_\alpha Q_{\mu\nu} \frac{\delta {\cal F}}{ \delta \partial_\beta Q_{\mu\nu}} =& -\kappa_0 \partial_\alpha Q_{\mu\beta} \partial_\nu Q_{\mu\nu}\\
&-\kappa_1 \partial_\alpha Q_{\mu\nu} \left( \partial_\beta Q_{\mu\nu} - \partial_\mu Q_{\nu\beta}\right).
\label{delgradq_stress}
\end{split}
\end{equation}
\rev{The hydrodynamic stress tensor is defined as
\begin{equation}
\Pi_{\alpha\beta}^{(HD)}=-p\,\delta_{\alpha\beta}-\rho u_\alpha u_\beta +
\eta(\partial_\beta u_\alpha + \partial_\alpha u_\beta) + \zeta \partial_\mu u_\mu \delta_{\alpha\beta},
\label{equation-hydro-stress}
\end{equation}}
where $\eta$ and $\zeta$ are the dynamic and bulk viscosity, respectively. The pressure $p$ is related to the density via an ideal gas equation of state as $p=c_s^2\rho$ with $c_s$ as lattice speed of sound as is standard in lattice Boltzmann. The last term vanishes in incompressible fluids as Eq. \ref{eq_mass1} becomes $\partial_\alpha u_\alpha=0$. No-slip and no-penetration boundary conditions are applied on the walls and particle surfaces, and 
the boundary conditions for $\bm{Q}$ are found from the minimisation of the free energy \cite{Skarabot2007}
\begin{equation}
\rev{n_\gamma} \frac{\partial \rev{f}}{\partial Q_{\alpha\beta,\gamma}}
+ \frac{\partial \rev{f_s}}{\partial Q_{\alpha\beta}} = 0,
\label{equation-lc-general-bc}
\end{equation}
where $Q_{\alpha\beta,\gamma}=\partial Q_{\alpha\beta}/\partial x_\gamma$.

\rev{The total force on the particle consists of a thermotropic contribution $\bm{F}^{(LC)}$ and a hydrodynamic contribution $\bm{F}^{(HD)}$. The thermotropic contribution is the integral of the gradient of the stress tensor $\bm{\Pi}^{(LC)}$ in Eq.~\ref{equation-lc-stress} over the surface $S$ of the particle. This can be separated into contributions from the bulk and gradient free energy:
\begin{eqnarray}
F^{(LC)}_\alpha &=& \int_S \partial_\beta \Pi_{\alpha\beta}\, dS\nonumber\\
&=& \int_S \partial_\beta \left(\Pi^{(b)}_{\alpha\beta}+\Pi^{(g)}_{\alpha\beta}\right)\, dS\nonumber\\
&=& F_\alpha^{(b)} + F_\alpha^{(g)}
\label{thermotropic_force}
\end{eqnarray}
Splitting the molecular field $\bm{H}$ in Eq.~\ref{eq-lc-h-full} into terms that contain only bulk and gradient contributions,
\begin{eqnarray}
\hspace*{-0.3cm}H^{(b)}_{\alpha\beta} &=& -A_0 (1 - \gamma/3) Q_{\alpha\beta} + A_0 \gamma\, (Q_{\alpha\mu} Q_{\mu\beta} - {\textstyle\frac{1}{3}} Q_{\mu\nu}^2\delta_{\alpha\beta}),\nonumber\\
&&-A_0 \gamma\, Q_{\mu\nu}^2 Q_{\alpha\beta}\\
\hspace*{-0.3cm}H^{(g)}_{\alpha\beta} &=&\kappa_0 \partial_\alpha \partial_\mu Q_{\mu\beta} + \kappa_1 \partial_\mu(\partial_\mu Q_{\alpha\beta} - \partial_\alpha Q_{\mu\beta}),
\end{eqnarray}
we obtain together with Eqs.~\ref{sym_stress}, \ref{antisym_stress} and \ref{delgradq_stress} for the
bulk contribution 
\begin{eqnarray}
\hspace*{-0.3cm}\Pi^{(b)}_{\alpha\beta} &=& f_b -\xi H^{(b)}_{\alpha\mu}(Q_{\mu\beta} + {\textstyle\frac{1}{3}}\delta_{\mu\beta})- \xi (Q_{\alpha\mu} + {\textstyle\frac{1}{3}}\delta_{\alpha\mu}) H^{(b)}_{\mu\beta}\nonumber\\
&&+ 2\xi (Q_{\alpha\beta} + {\textstyle \frac{1}{3}}\delta_{\alpha\beta}) Q_{\mu\nu} H^{(b)}_{\mu\nu} +Q_{\alpha\mu} H^{(b)}_{\mu\beta} - H^{(b)}_{\alpha\mu} Q_{\mu\beta}\nonumber\\
\end{eqnarray}
and for the gradient contribution
\begin{eqnarray}
\hspace*{-0.3cm}\Pi^{(g)}_{\alpha\beta} &=& f_g -\xi H^{(g)}_{\alpha\mu}(Q_{\mu\beta} + {\textstyle\frac{1}{3}}\delta_{\mu\beta})- \xi (Q_{\alpha\mu} + {\textstyle\frac{1}{3}}\delta_{\alpha\mu}) H^{(g)}_{\mu\beta}\nonumber\\
&&+ 2\xi (Q_{\alpha\beta} + {\textstyle \frac{1}{3}}\delta_{\alpha\beta}) Q_{\mu\nu} H^{(g)}_{\mu\nu} +Q_{\alpha\mu} H^{(g)}_{\mu\beta} - H^{(g)}_{\alpha\mu} Q_{\mu\beta}\nonumber\\
&& +\kappa_0 \partial_\alpha Q_{\mu\beta} \partial_\nu Q_{\mu\nu} + \kappa_1 \partial_\alpha Q_{\mu\nu} \left( \partial_\beta Q_{\mu\nu} - \partial_\mu Q_{\nu\beta}\right).
\end{eqnarray}
The hydrodynamic contribution to the force on the particle is also given through the surface integral of the gradient of the hydrodynamic stress tensor $\bm{\Pi}^{(HD)}$ over the particle surface $S$ as
\begin{equation}
F^{(HD)}_\alpha =\int_S \partial_\beta \Pi_{\alpha\beta}^{(HD)} dS.
\label{hydro_force}
\end{equation}
See the following section \ref{simulation_method} on how these forces are evaluated.
}

\section{Simulation Method}\label{simulation_method}

\subsection{Simulation Setup}
\begin{figure}[htbp]
\centering
\begin{tikzpicture}[scale=0.9]

    \fill[gray!50] (0,4,0) -- (0,4,2) -- (8,4,2) -- (8,4,0) -- cycle;
    \fill[gray!50] (0,0,0) -- (0,0,2) -- (8,0,2) -- (8,0,0) -- cycle;

	\draw[very thick] (0,0,0) rectangle (8,4,0);
	\draw[very thick] (0,0,2) rectangle (8,4,2);
	
	\draw[very thick] (0,0,0) -- (0,0,2);
	\draw[very thick] (0,4,0) -- (0,4,2);
	\draw[very thick] (8,0,0) -- (8,0,2);
	\draw[very thick] (8,4,0) -- (8,4,2);
	
    \path[draw,thick] (0,0,1) .. controls (4,1.334 ,1) and (4,2.666,1) .. (0,4,1);
    \draw [-stealth](0,2,1) -- (3,2,1);
    \draw [-stealth](0,1,1) -- (2.1,1,1);
    \draw [-stealth](0,3,1) -- (2.1,3,1);
    \draw [-stealth](0,1.5,1) -- (2.7,1.5,1);
    \draw [-stealth](0,2.5,1) -- (2.7,2.5,1);
    \draw [-stealth](0,0.5,1) -- (1.2,0.5,1);
    \draw [-stealth](0,3.5,1) -- (1.2,3.5,1);
    \shade[ball color = lightgray, opacity = 0.75] (4.7,2,1) circle (0.5cm);
    
    \node[] at (3.8,-1.30) {\small No-slip, no-penetration, homeotropic anchoring };
    \node[] at (3.8, 4.50) {\small No-slip, no-penetration, homeotropic anchoring };
    \node[] at (2.8,0.8) {\small fluid};
    \node[] at (2.8,0.4) {\small velocity $\bm{u}$};
    \node[] at (4.8,2.6) {\small No-slip, no-penetration,};
    \node[] at (5.85,2.2) {\small homeotropic};
    \node[] at (6.1,1.8) {\small anchoring};
    \node[] at (4.9,0.9) {\small radius $R$ };
    
    \node[] at (8.4,2) {\large{$L_x$} };
    \draw[->, thick] (8.4,2.3)  -- (8.4,4) ;
    \draw[->, thick] (8.4,1.7)  -- (8.4,0); 
    
    \draw[thick,->] (-0.40,-0.36) -- (0.75,-0.36);
    \node[] at (1,-0.36) {$z$ };
    \draw[thick,->] (-0.40,-0.36) -- (-1,-1.05);
    \node[] at (-1.1,-1.3) {$y$ };
    \draw[thick,->] (-0.40,-0.36) -- (-0.4,0.88);
    \node[] at (-0.6,0.9) {$x$ };

\end{tikzpicture}
\caption{Sketch of the computational geometry: We apply no-slip boundary conditions and homeotropic anchoring conditions at the walls in $x$-direction and periodic boundary conditions at the $y$- and $z$-boundaries.}
\label{setup}
\end{figure}
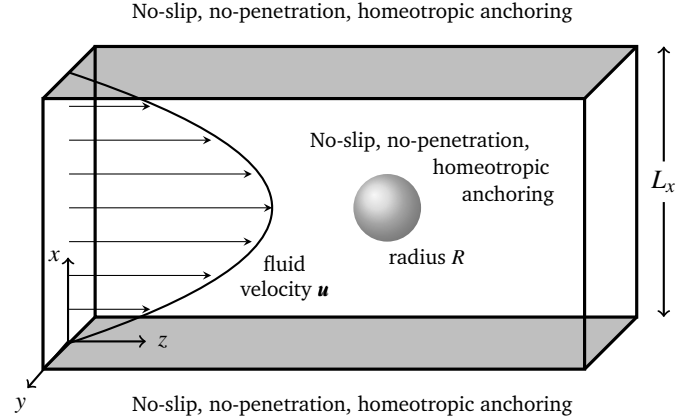

Fig~\ref{setup} shows a sketch of the three dimensional computational geometry, which consists of a channel of dimensions $L_x \times L_y \times L_z = 128 \times 64 \times 256$ lattice sites. 

\rev{We used a hybrid lattice Boltzmann scheme \cite{Marenduzzo2007} that treats the dynamics of the $\bm{Q}$-tensor order parameter with a finite-difference scheme and applies the lattice Boltzmann method to the hydrodynamic variables.}
Parallel, solid walls are positioned at $x=0$ and $x=L_x$, whereas periodic boundary conditions are applied in $y$- and $z$-direction with the $z$-boundaries acting as inlet and outlet of the microchannel. 
The pressure gradient $\Psi=\Delta p/L_z$ is applied in $z$-direction as body force density that acts on all fluid sites \rev{besides the forces that arise from the thermotropic and hydrodynamic stresses in Eqs~\ref{equation-lc-stress} and \ref{equation-hydro-stress}, respectively. Our approach uses Guo forcing \cite{Guo2002, LBBook2017} to apply all forces on the fluid, which removes undesirable artefacts that may enter the continuity and momentum equation due to the time discretisation.}
 
The colloidal particles are discretised as solid, mobile particles with a radius of $R=7.2$ or $9.6$ lattice sites. All of our results use the radius $R=9.6$ apart from those presented in Fig.~\ref{trajectories_7_2}. As mentioned above, on the walls and particle surfaces no-slip and no-penetration boundary conditions are imposed through the bounce-back on links scheme \cite{Ladd:1994a, LBBook2017}. The surface free energy in Eq.~\ref{surface_fe} invokes a weak homeotropic anchoring condition with a preferred orientation of the director normal to the surfaces. 
\rev{The thermotropic force on the particle $\bm{F}^{(LC)}$ in Eq.~\ref{thermotropic_force} is integrated using a finite-difference scheme in $\bm{Q}$ and its gradients. The hydrodynamic force $\bm{F}^{(HD)}$ in Eq.~\ref{hydro_force} is integrated in a similar way using the hydrodynamic stress tensor $\bm{\Pi}^{(HD)}$, which is directly accessible in the lattice Boltzmann method through second order moments of the non-equilibrium distributions. The total force $\bm{F}=\bm{F}^{(LC)}+\bm{F}^{(HD)}$ is fed into a molecular dynamics algorithm to integrate the motion of the particles.}    

\rev{Several technical limitations of our model should be noted. While the centre of mass of the particle is integrated off-grid according to Newton's equation, the particle itself is discretised using a stair-case geometry. This requires remapping of the particle onto the lattice as the particle moves. Consequently, this can entail spikes in the force at single iteration steps, although these quickly average out over a few iteration steps with no detrimental effect on the trajectories. The pressure obeys an ideal gas equation of state and is directly related to the density via $p=c_s^2\rho$, while the effect of the constant pressure gradient is modelled through an additional body force density on the fluid. Both treatments are common in the lattice Boltzmann methodology and allow for an accurate modelling of a weakly compressible fluid, but the assumption of a constant pressure gradient represents obviously a simplification over the real situation. Thermal fluctuation have not been included as our simulations were carried out at a point in the phase diagram that is deep in the ordered state well away from the isotropic-nematic transition line where elastic forces from the anchoring of the liquid crystal dominate over thermal forces by orders of magnitude. Our approach uses the one-constant approximation to simplify elasticity where the elastic constants for splay, bend and twist deformations carry the same value. This approximation is commonly used as first approach and does not compromise our results qualitatively. However, relaxing this approximation could lead to quantitative differences, and potentially also richer phenomenology.  
The Beris-Edwards model uses a simplified approach to viscosities compared to the Ericksen-Leslie theory, which has six viscosity coefficients $\alpha_1,\dots,\alpha_6$ \cite{deGennes} (only five are independent as the Parodi relation applies \cite{Parodi1970}). The viscosities in the Beris-Edwards model are implicitly given through the isotropic dynamic shear viscosities $\eta$, the rotational diffusion constant $\Gamma$, the flow alignment parameter $\xi$ and the scalar order parameter $S_0$. They can be directly related to the Ericksen-Leslie viscosities $\alpha_1,\dots,\alpha_6$ \cite{Marenduzzo2007}, but parameterise only a subset of possible values.
}

\rev{All simulations were run with {\it Ludwig}, our lattice Boltzmann code for complex fluids, version 0.15.0.} Typical simulations ran for $8\times10^5$ iteration steps at various pressure gradients, each of which took approximately 16 hours to complete using a hybrid Message Passing Interface/OpenMP parallelisation with 4 MPI-tasks each running on 20 OpenMP threads. Model parameters used for the simulations is provided in Table \ref{parameters}.
\rev{More information about the specific implementation used in this work can be also found in the {\it Ludwig} code repository \cite{Ludwig} and related literature \cite{Desplat2001, Adhikari2005}.}
\begin{table}[htb]
\normalsize
\centering
\begin{tabular}{| l | c | c |}
\hline
Bulk energy scale & $A_0$ & 0.01\\
\hline
Inverse temperature & $\gamma$ & 3.1\\
\hline
Elastic constants & $\kappa_0, \kappa_1$ & 0.01\\
\hline
Wall anchoring strength & $w_{wall}$ & 0.02\\
\hline
Particle anchoring strength & $w_{part}$ & 0.01\\
\hline
Flow alignment parameter & $\xi$ & 0.7\\
\hline
Mobility parameter& $\Gamma$ & 0.5\\
\hline
\rev{Density} & \rev{$\rho$} & \rev{1.0}\\
\hline
Dynamic viscosity & $\eta$ & 5/6\\
\hline
Bulk viscosity & $\zeta$ & 5/6\\
\hline
Particle radius & $R$ & 7.2, 9.6\\
\hline
\end{tabular}
\caption{Overview of simulation parameters}
\label{parameters}
\end{table}

\subsection{Parameter Mapping to Physical Units}

In order to map our simulation units to physical units we need to calibrate the units of length, time and pressure. To this end, we relate the lattice spacing $\Delta x$, the algorithmic time step $\Delta t$ and the reference pressure $p^*$, which are all unity in lattice Boltzmann units (LBU), to their values in SI units.

The calibration of the length scale is straightforward as it is simply set by considering the diameter $D$ of the colloidal particle. Assuming the largest radius particle that we consider corresponds to a relatively small diameter of $D=0.2\mu$m in SI units, results in a LBU of length $\Delta x \,\widehat{=}\, 10^{-8}$ m $= 10$ nm in SI units. This length scale allows for an accurate resolution of the liquid-crystalline order structure and flow field around the particle, while keeping the necessary computational resource relatively low.

To obtain a pressure scale, we use the measurements of the Landau-de Gennes parameters \cite{WrightMermin:1989} (see Appendix D therein), which suggest
\begin{eqnarray*}
\frac{27}{2\,A_0\,\gamma}&\simeq& 5\times 10^{-6} \mbox{ J}^{-1} \mbox{m}^3=5\times 10^{-6} \mbox{ Pa}^{-1}
\end{eqnarray*}
for some liquid crystals. Using $A_0=0.01$ and $\gamma=3.1$ in our simulations leads to a reference pressure of $p^* = 1$ LBU $\widehat{=}\, 10^8$ Pa in SI units.

For the timescale calibration we use the following formula, which relates the rotational viscosity $\gamma_1$ of the director, to the equilibrium scalar order parameter $q$ and the order parameter mobility $\Gamma$:
\begin{equation*}
   \gamma_1 = \frac{2 q^2}{\Gamma}
\end{equation*}
We use $\Gamma= 0.5$ and bulk energy density parameters that give $q \approx 0.5$ since it is assumed that the system is well within the nematic phase. Therefore, the rotational viscosity $\gamma_1=1$ LBU. Typical values for liquid crystals in SI units are $\gamma_1=0.1$ Pa s \cite{deGennes}. Together with 1 Pa equating to a pressure of $10^{-8}$ in LBU, we obtain for the algorithmic time step $\Delta t\,\widehat{=}\,10^{-9}$ s $= 1$ ns.

The Reynolds number gives the ratio of inertial to viscous forces, and in our approach this can be estimated on the basis of typical velocities of colloidal particles as they are advected with the flow. For instance, in a later simulation (Fig. \ref{trajectories}) with $\Psi=9.6\times10^{-6}$ and $R=9.6$, we observe that a colloidal particle that migrates preferentially to the attractor region at $x\simeq 47$ travels around $1.6\times10^4 \Delta x$ in the $z$-direction during the $8\times10^5 \Delta t$ (all simulations in Figs. \ref{noLC}, \ref{trajectories} and \ref{trajectories_7_2} were run for $8\times10^5$ algorithmic steps). This leads to a velocity $v\simeq 0.02$ in LBU. With the values for density $\rho$ and dynamic viscosity $\eta$ as specified in   Table \ref{parameters} and a diameter of a colloidal particle $D=20 \Delta x$ as typical length scale $\Lambda$, this results in a Reynolds number 
\begin{equation*}
Re = \frac{\rho\, v \, \Lambda}{\eta} \simeq 0.48
\end{equation*}
for this pressure difference, and indicates a flow regime \rev{where viscous forces are larger than, but comparable to inertial forces}.

The Ericksen number gives a ratio of viscous to elastic forces and, with the above values of the flow velocity $v$, dynamic viscosity $\eta$, characteristic length scale $\Lambda$ and either elastic constant $\kappa_0$ or $\kappa_1$ (see   Table \ref{parameters}), we obtain
\begin{equation*}
Er = \frac{\eta\,v\,\Lambda}{\kappa} \simeq 33.33,
\end{equation*} 
which defines a flow regime where the director field is strongly influenced by the flow.

\section{Results and Discussion}

Before addressing preferential particle migration in nematic host phases, we begin with presenting some general results for contextualisation and later reference. 

\begin{figure}[htbp!]
\centering
\includegraphics[trim={0cm 0.5cm 0cm 0cm},clip,scale=0.35]{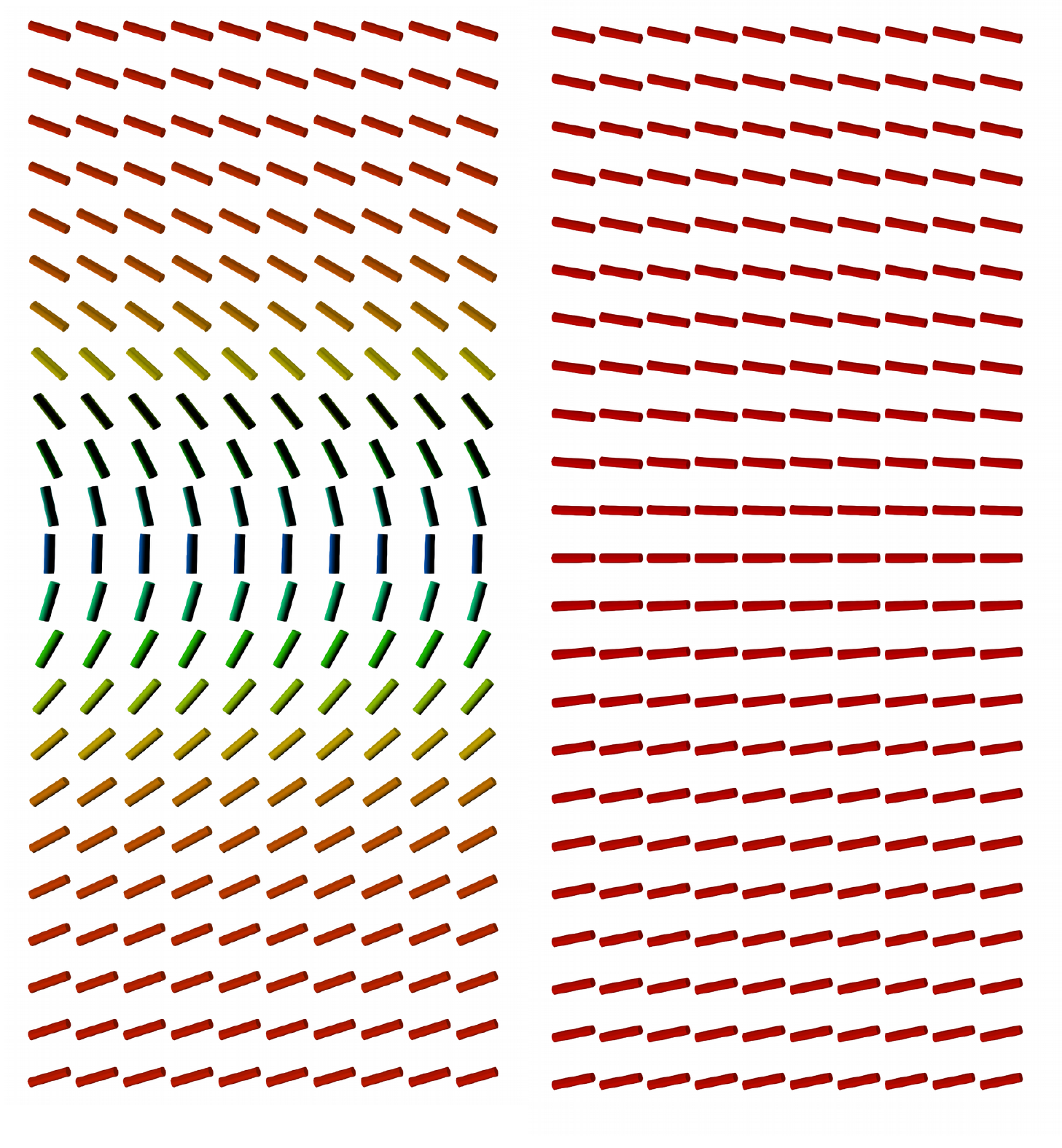}
\caption{Bend or H-state (left) and splay or V-state (right) at the centre of the channel.}
\label{bendsplaystate}
\end{figure}

In the quiescent state the homeotropic wall anchoring induces a nematic order with a director orientation parallel to the wall normals ($z$-direction in Fig. \ref{setup}). When the flow is switched on, flow-alignment of the director to the appropriate Leslie angle takes place. 
The nematic liquid crystal can adopt two possible conformations, bend or splay, as shown in Fig.~\ref{bendsplaystate}.  

In both states the director flow-aligns to a positive Leslie angle in the lower half of the channel (where there is positive shear) and  to a negative Leslie angle in the upper half of the channel (where there is negative shear).
The bend state (sometimes called the H-state) and splay state (sometimes called the V-state) differ in how the director rotates between the positive and negative Leslie angles at the centre of the channel. In the bend state the director at the centre is perpendicular to the walls and in the splay state the director at the centre is parallel to the walls.  

\begin{figure}[htbp]
\centering
\includegraphics[trim={2.0cm 0cm 0cm 0cm},clip,width=1.1\linewidth]{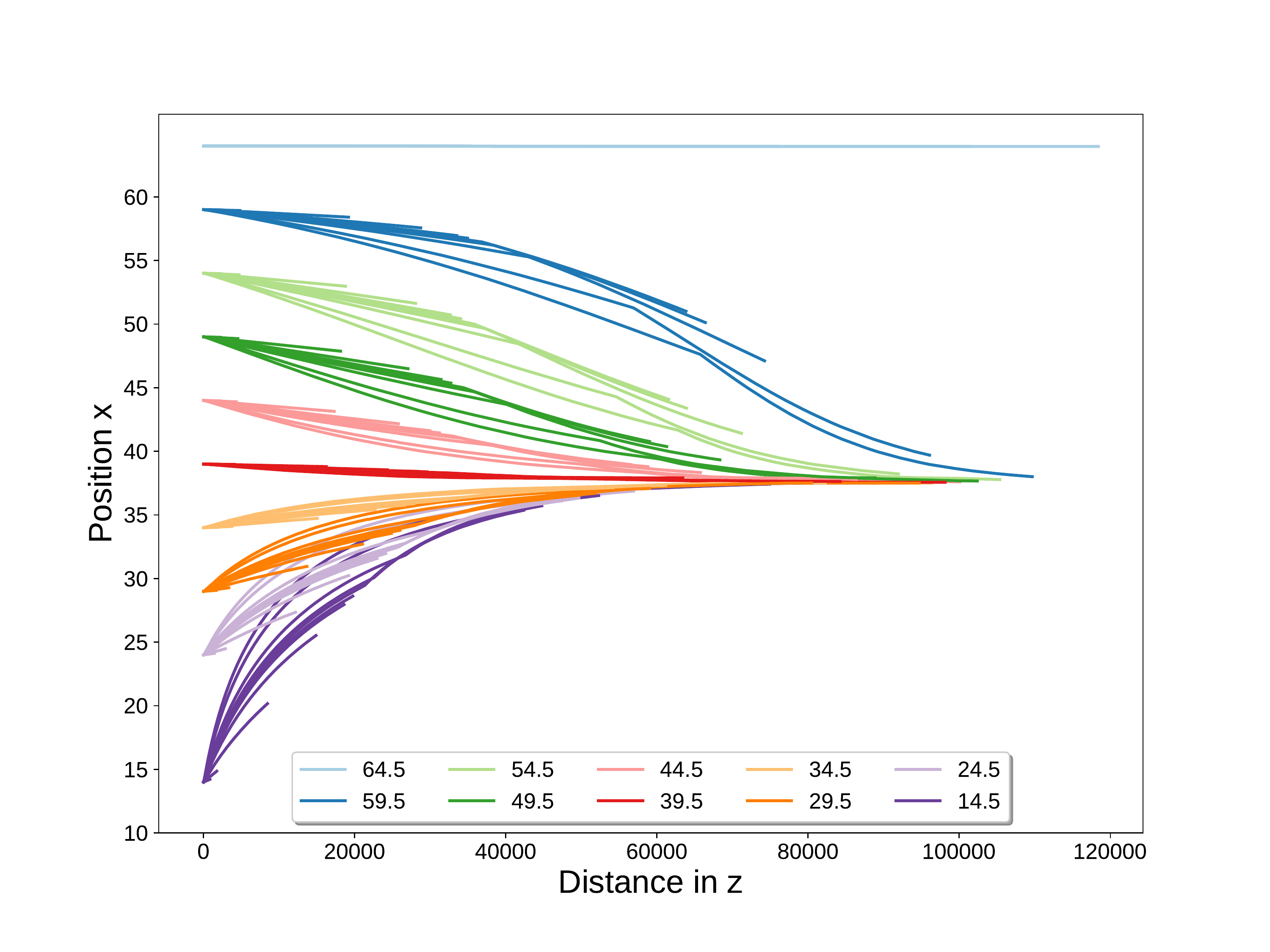} 
\caption{Particle migration in an isotropic fluid due to the Segr\'e-Silberberg effect.  Shown are particle position $x$, across the channel gap, versus the particle position $z$, along the channel, for a variety of starting positions (color coded) and pressure gradients in the range $1.25\times10^{-6} \le \Psi \le 1.75\times10^{-5}$. All quantities are given in LBU.}
\label{noLC}
\end{figure}

In the absence of any liquid crystalline order, i.e.~either in a classical Newtonian fluid or in a liquid crystal host phase at temperatures above the isotropic-nematic transition point, preferential migration of colloidal particles occurs to an $x$-position between the centre and wall of the channel. This is referred to as the Segr\'e-Silberberg effect and shown in Fig. \ref{noLC} through trajectories for all starting positions $x_s$ at all applied pressure gradients $\Psi$. Two aspects of the Segr\'e-Silberberg effect are important to understand when comparing to our results further below. Firstly, except for particles that start exactly at the centre of the channel, migration is to a single equilibrium $x$-position over time. Secondly, the value of $x$  does not vary significantly at low Reynolds numbers. As seen in Fig.~\ref{noLC}, in simulations we see exactly these aspects of the Segr\'e-Silberberg effect: all particles migrate to an equilibrium at $x$-position of approximately $38$ regardless of their starting position or applied pressure.

It should be noted that the Segr\'e-Silberberg equilibrium position seen in Fig.~\ref{noLC} is not at the well-known $0.6$ tube radii (or channel half-widths in the planar case) from the centre line, but at approximately $0.4$ channel half-widths.
This shift of the equilibrium position towards the channel centre has been observed before in simulations \cite{Kilimnik2011, Lashgari2017} and can be attributed to our particle confinement ratio of $R/L_x=19.2/128=0.15$. In contrast, the analytical results \cite{Ho1974, Schonberg1989} were obtained for zero confinement ratio, i.e.~for particles that are negligibly small compared to the tube diameter or gap width. Our study uses Reynolds numbers that are around two orders of magnitude smaller than those in previous simulation studies \cite{Kilimnik2011, Lashgari2017}, which is known to result in equilibrium positions that are again closer to the channel centre.

Using the model described in Section \ref{theory} we studied the migration behaviour of a single particle at different pressure gradients $\Psi (=\Delta p/L_z)$ and start positions $x_s$.

\begin{figure}[htbp!]
\centering
\includegraphics[trim={2.75cm 6.9cm 4.3cm 8.9cm},clip=true,width=\linewidth]{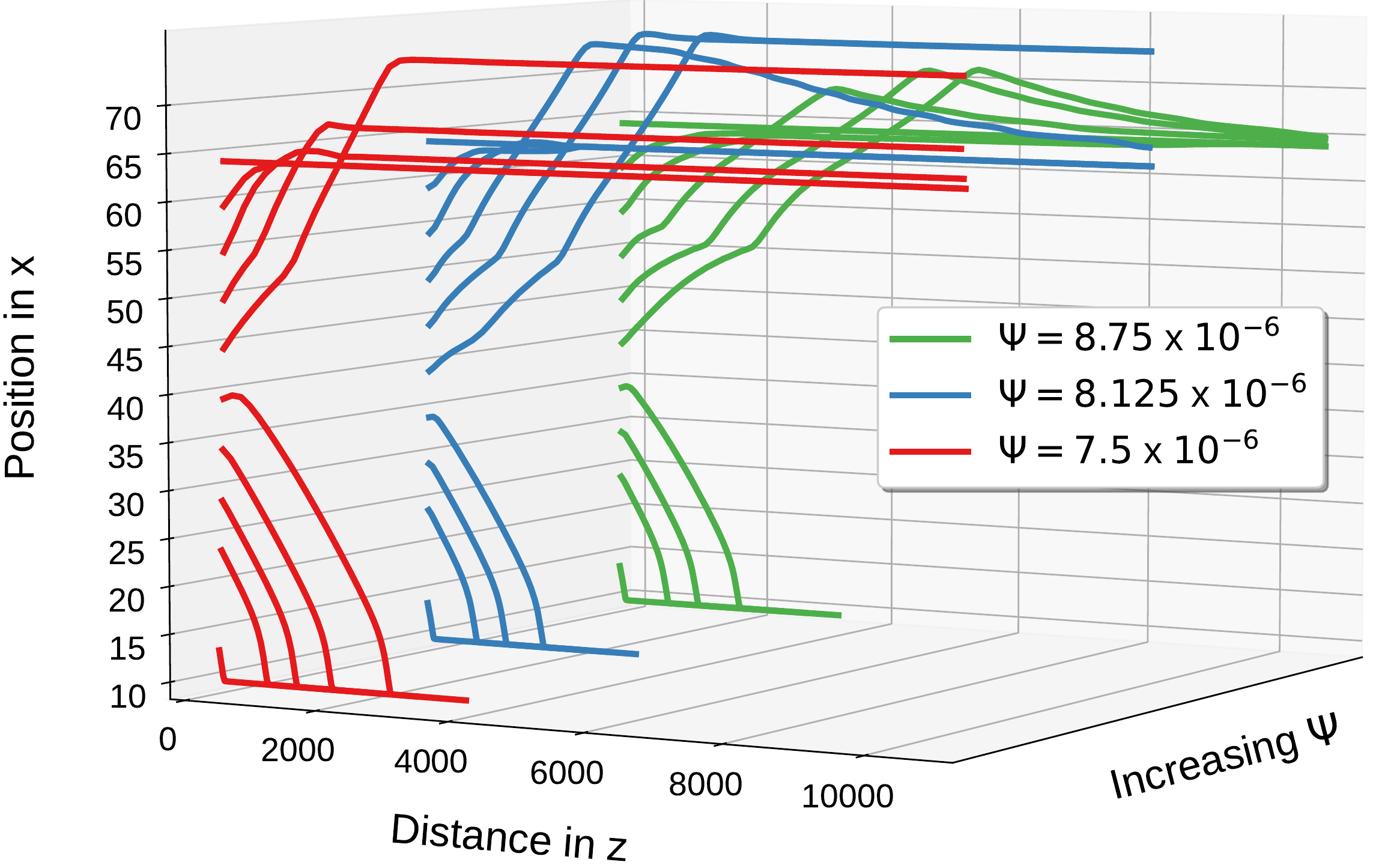} 
\caption{Particle trajectories for particle radius $R=9.6$ at sub-critical pressure gradients $\Psi=7.5\times 10^{-6},\,8.125\times 10^{-6},\,8.75\times 10^{-6}$, showing particle $x$-positions across the channel gap versus the $z$-distance travelled in the flow direction for various initial positions $x_s$. 
\rev{The particle Ericksen numbers $Er$ (particle Reynolds numbers $Re$) are from front to back $Er=25.62$ ($Re=0.37$), $Er=27.91$ ($Re=0.40$), $Er=30.20$ ($Re=0.43$). All quantities are given in LBU.}}
\label{sub_critical}
\end{figure}

Fig.~\ref{sub_critical} shows particle trajectories $x(z)$ for starting positions in  the lower portion of the channel (the channel centre is at $x=64$). For low pressure gradients $\Psi=7.5\times 10^{-6}$ (red lines in Fig.~\ref{sub_critical}) the particle migrates either to the channel walls (for starting position $x_s\le40$) or towards the centre of the channel (for $x_s\ge 44$). Particles that start at the centre stay in that position, while those inserted further away from the centre show a tendency to overshoot and remain in an off-centre position.
At slightly larger pressure gradients $\Psi = 8.125 \times 10^{-6},\,8.75\times 10^{-6}$ (blue  and green lines in Fig.~\ref{sub_critical}, respectively), this bi-stability remains, but with the division between migration towards the wall or centre now at around $x_s\simeq 38$. Some trajectories exhibit the onset of a pull-back behaviour from overshot off-centre positions towards the centre, in particular for the larger pressure gradient of $\Psi=8.75\times 10^{-6}$. 
It should be noted that for low pressure gradients, $\Psi\le5\times 10^{-6}$, we also observe off-centre equilibrium positions in form of a weak attractor that moves to the channel centre with increasing pressure gradient.

\begin{figure}[htbp!]
\centering
\includegraphics[trim={3.3cm 6.5cm 4.3cm 7.5cm},clip,width=\linewidth]{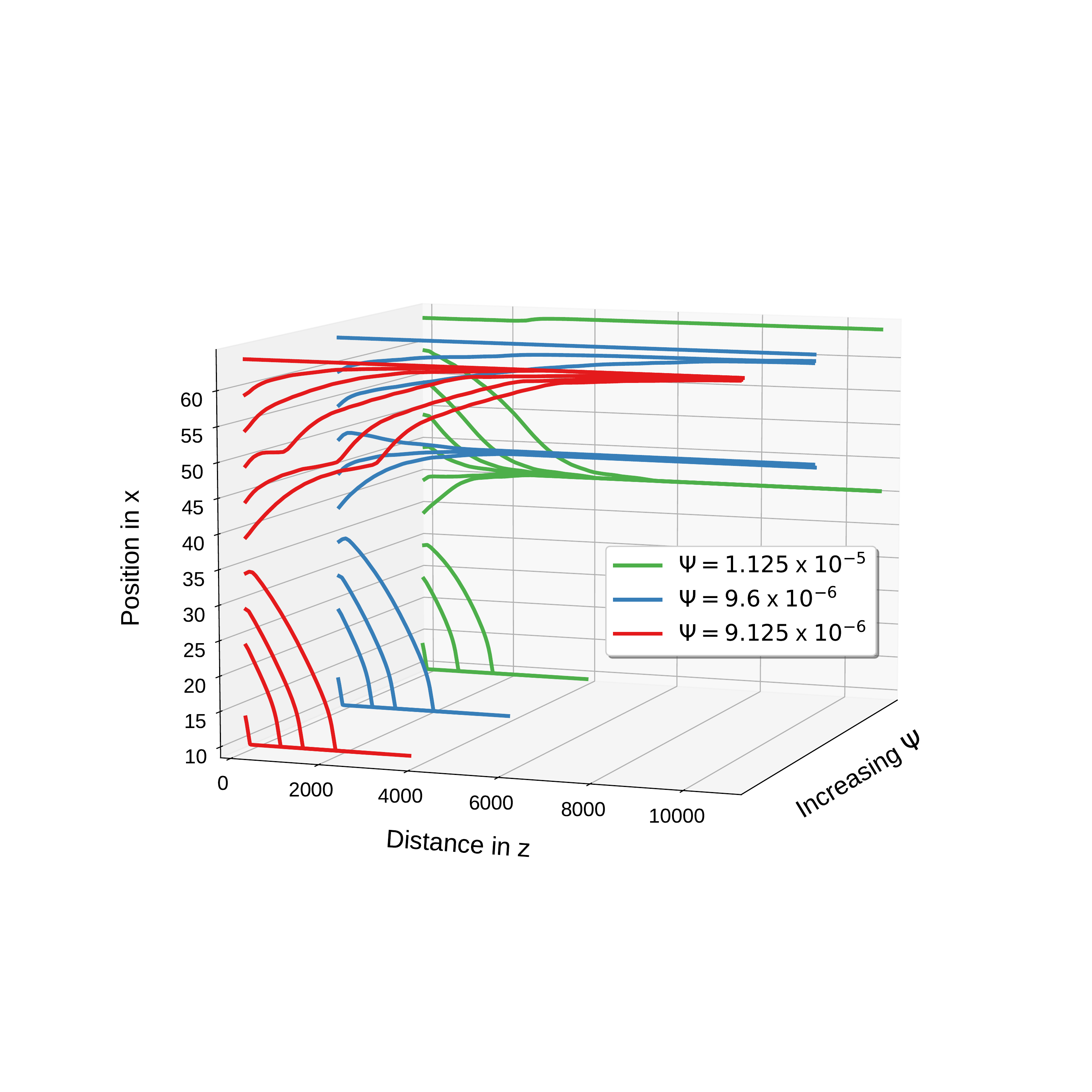} 
\caption{Particle trajectories for particle radius $R=9.6$ at pressure gradients $\Psi=9.125\times 10^{-6},\,9.6 \times 10^{-6},\,1.125 \times 10^{-5}$, showing particle $x$-positions across the channel gap versus the $z$-distance travelled in the flow direction for various initial positions $x_s$. The emergence of the particle attractor is shown for $\Psi=9.6 \times 10^{-6},\,1.125 \times 10^{-5}$. \rev{The particle Ericksen numbers $Er$ (particle Reynolds numbers $Re$) are from front to back $Er=31.58$ ($Re=0.45$), $Er=33.33$ ($Re0.48$), $Er=39.57$ ($Re0.57$). All quantities are given in LBU.}}
\label{attractor}
\end{figure}

A further increase in the pressure gradient leads to a sudden onset of a new type of preferential migration behaviour.
In Fig.~\ref{attractor} we see for $\Psi =9.125\times 10^{-6}$ that a pronounced trajectory kink emerges at $x\simeq 50-53$, the precursor of which is also visible for lower pressure gradients in Fig.~\ref{sub_critical}. For the pressure gradients $\Psi = 9.6\times 10^{-6}$ (blue lines in Fig.~\ref{attractor}) this kink transitions into a third, emergent particle attractor, in addition to the channel centre and wall. At $\Psi=9.6\times 10^{-6}$ this emergent attractor is located at $x \simeq 48$ and trajectories with initial positions $40<x_s<50$ migrate towards the attractor. Increasing the pressure gradient further, leads to a movement of the emergent attractor towards the wall, and for $\Psi = 1.125\times10^{-5}$ the attractor position is $x \simeq 40$.

\begin{figure*}[htbp!]
\centering
\begin{tabular}{ccc}
\includegraphics[trim={2cm 1.2cm 2.1cm 2.6cm},clip, scale=0.23]{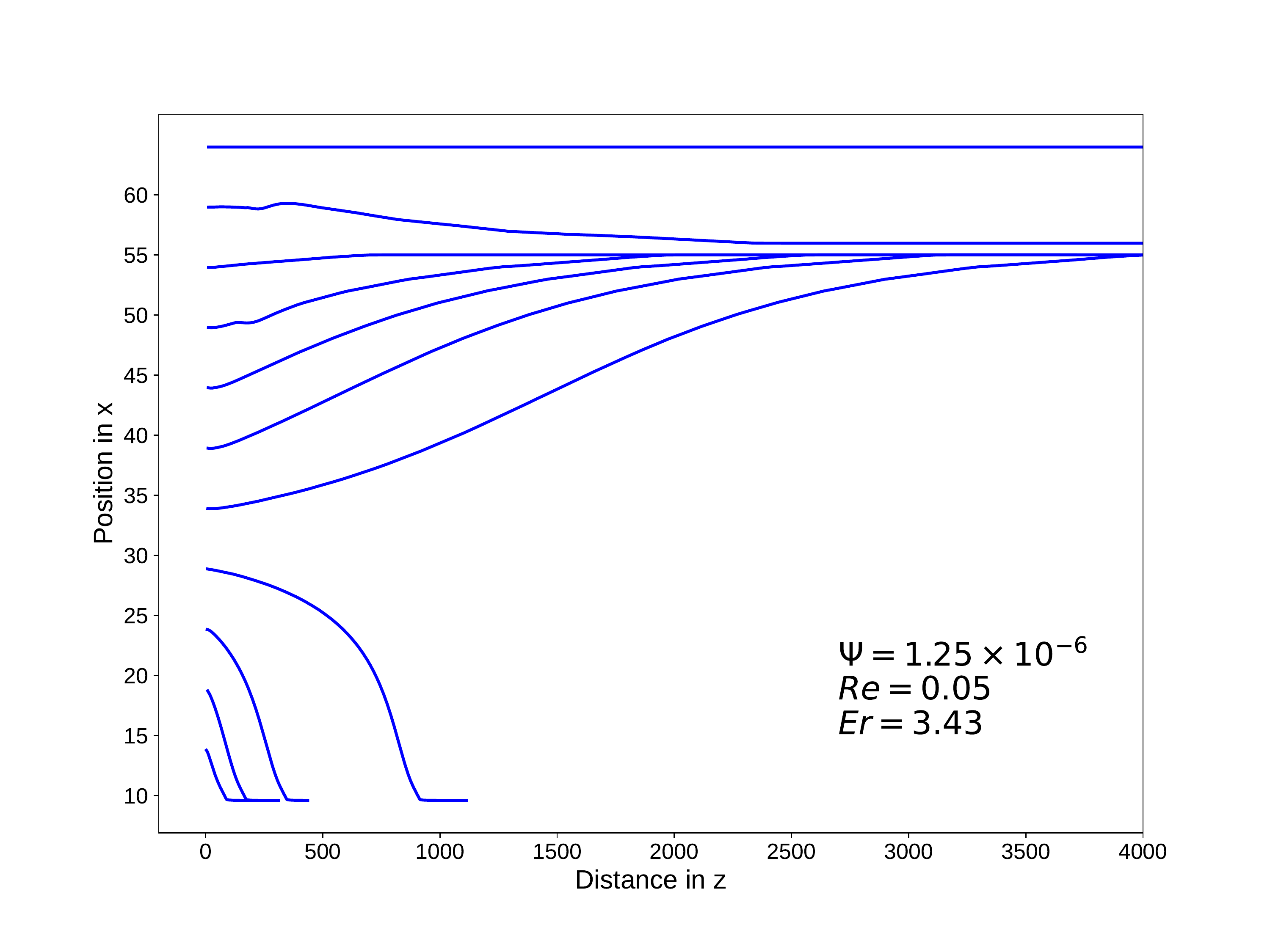}
\includegraphics[trim={2cm 1.2cm 2.1cm 2.6cm},clip, scale=0.23]{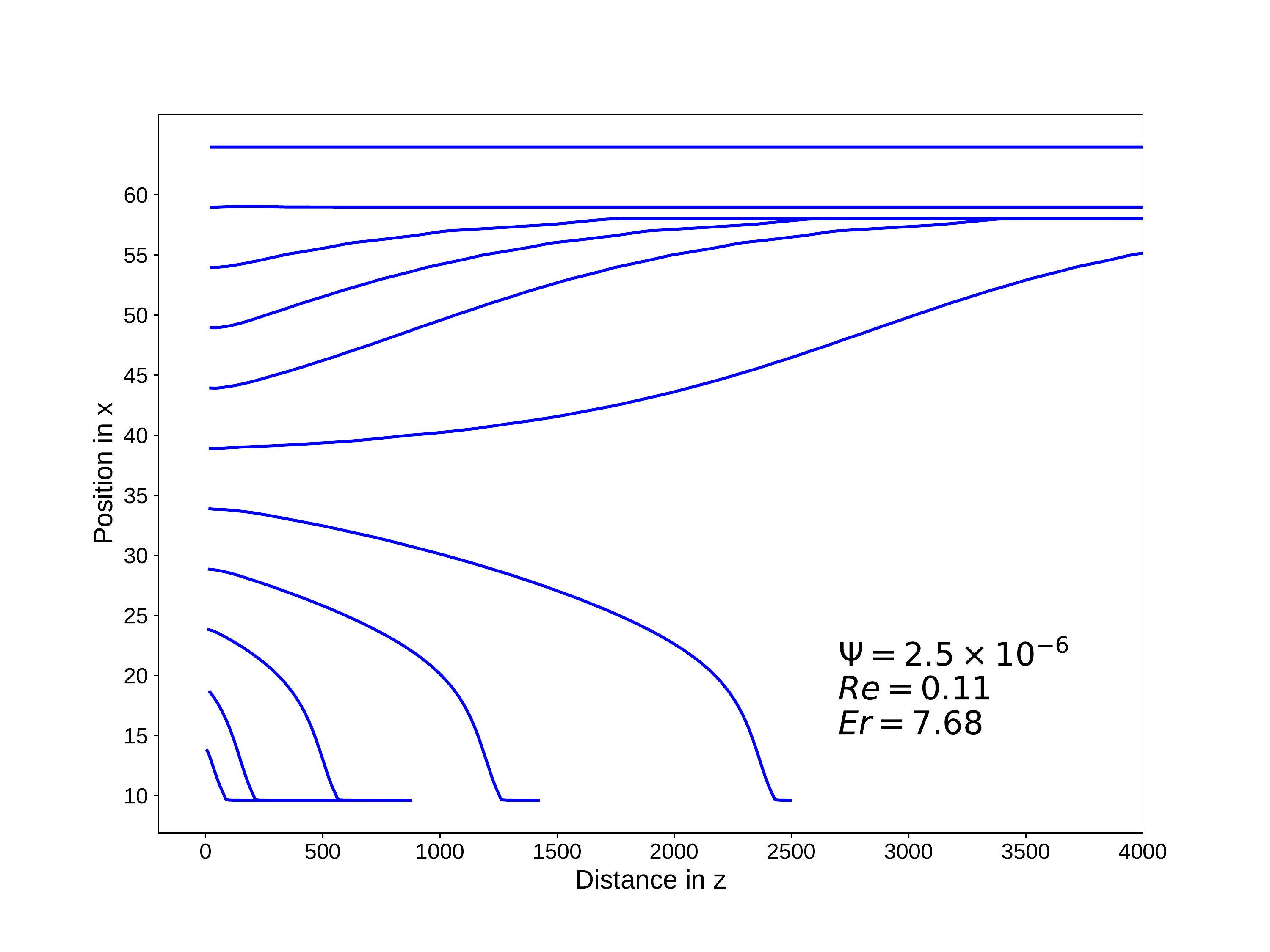}
\includegraphics[trim={2cm 1.2cm 2.1cm 2.6cm},clip, scale=0.23]{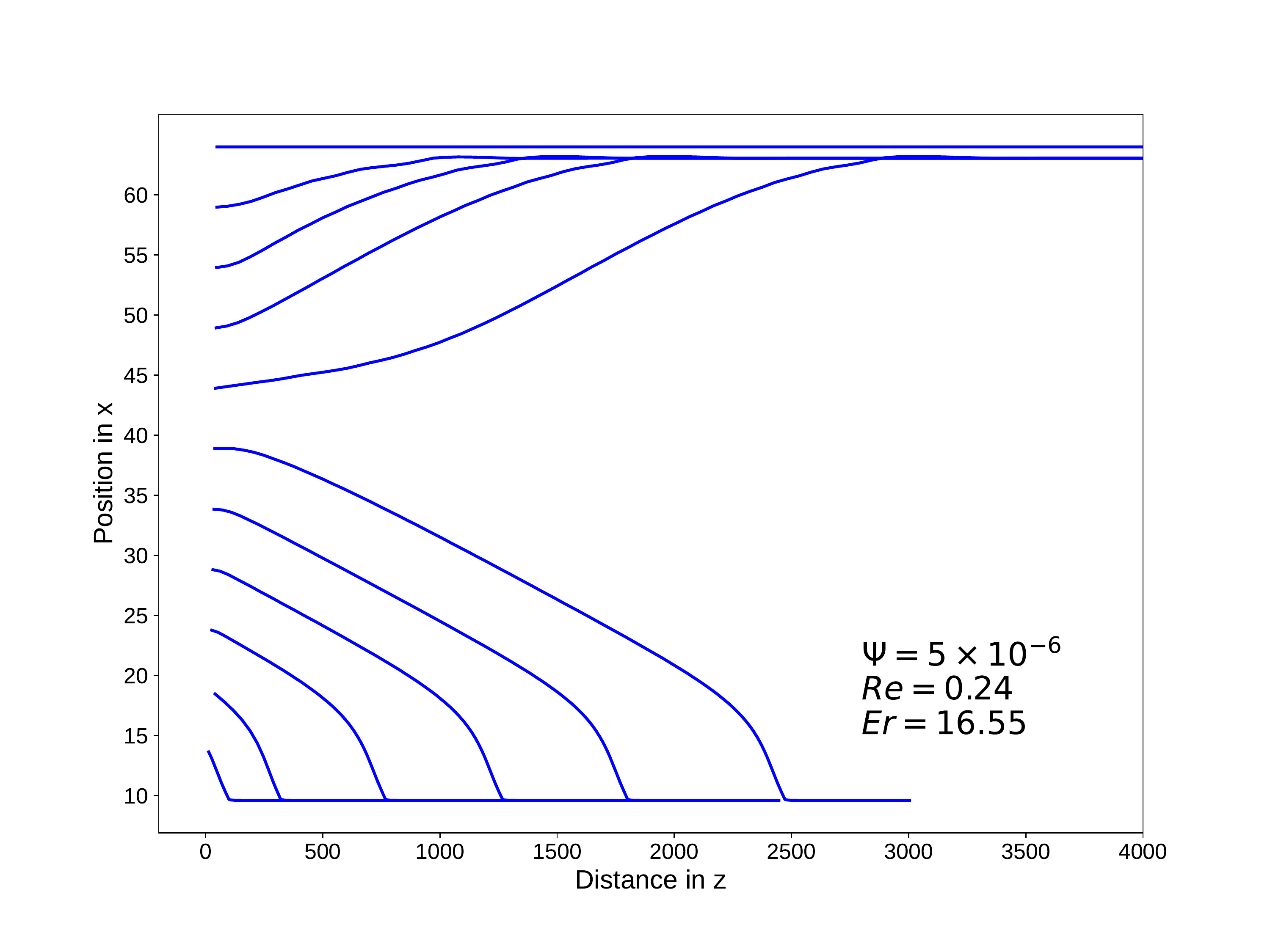} \\
\includegraphics[trim={2cm 1.2cm 2.1cm 2.6cm},clip, scale=0.23]{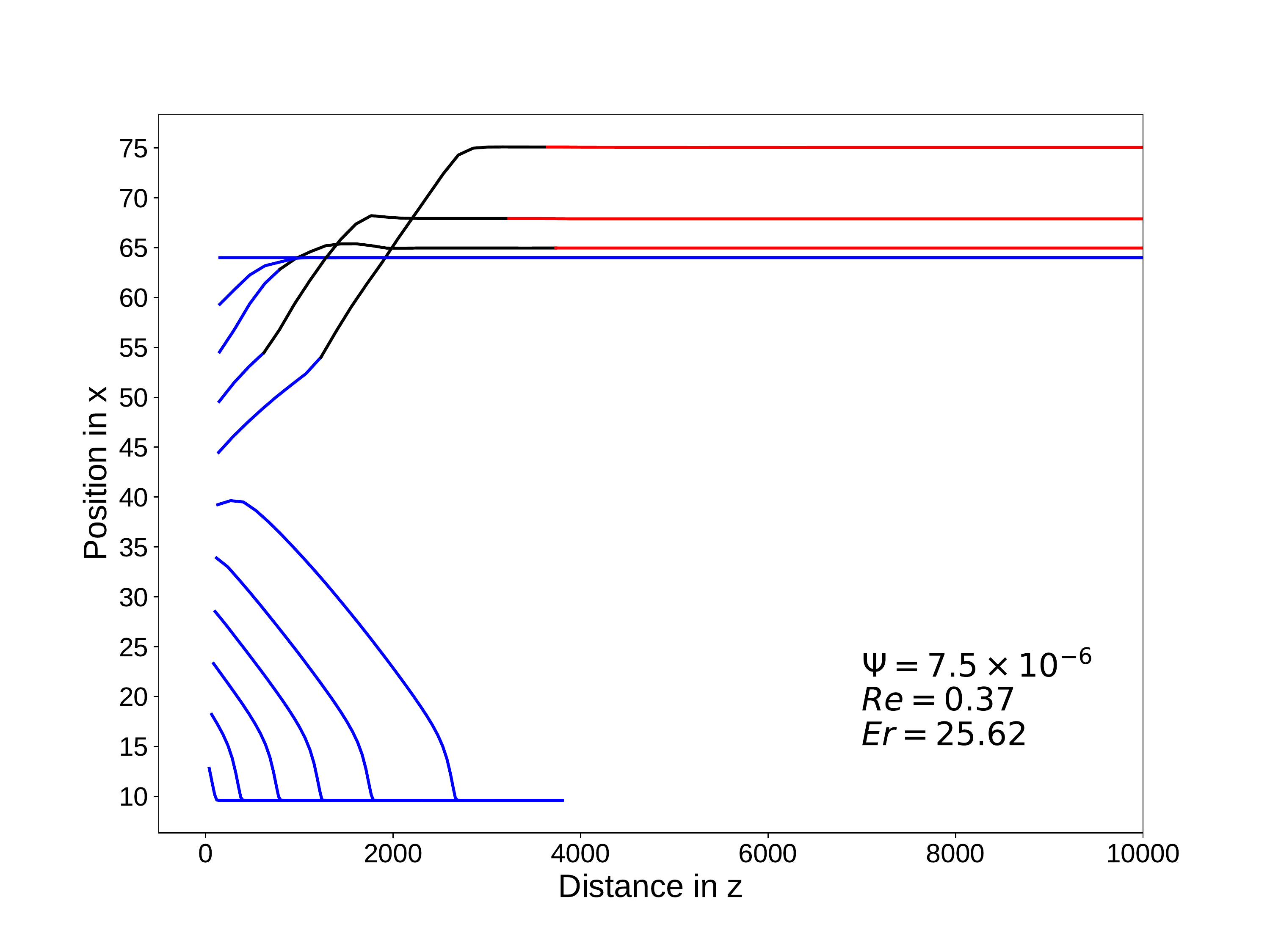}
\includegraphics[trim={2cm 1.2cm 2.1cm 2.6cm},clip, scale=0.23]{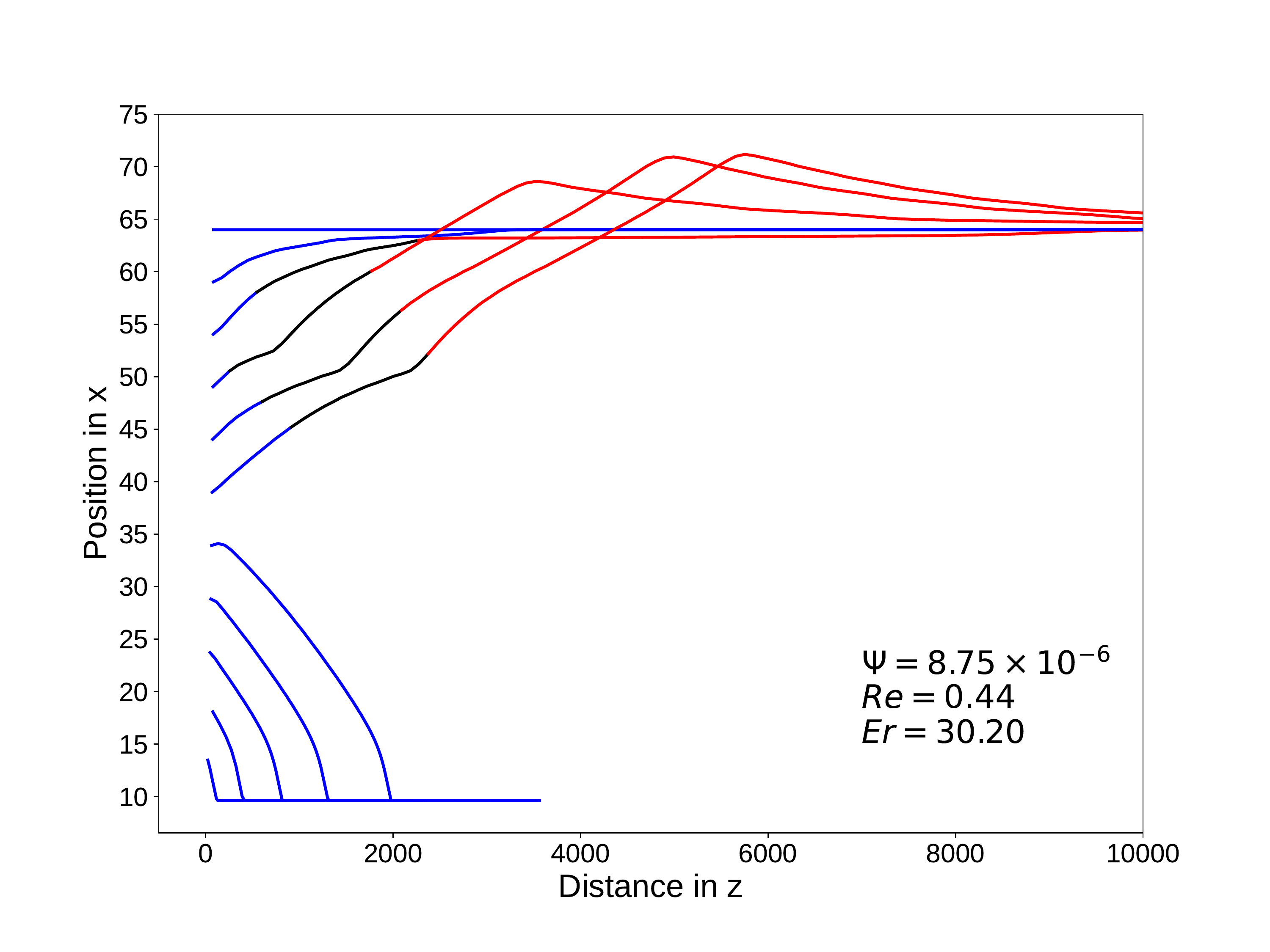}
\includegraphics[trim={2cm 1.2cm 2.1cm 2.6cm},clip, scale=0.23]{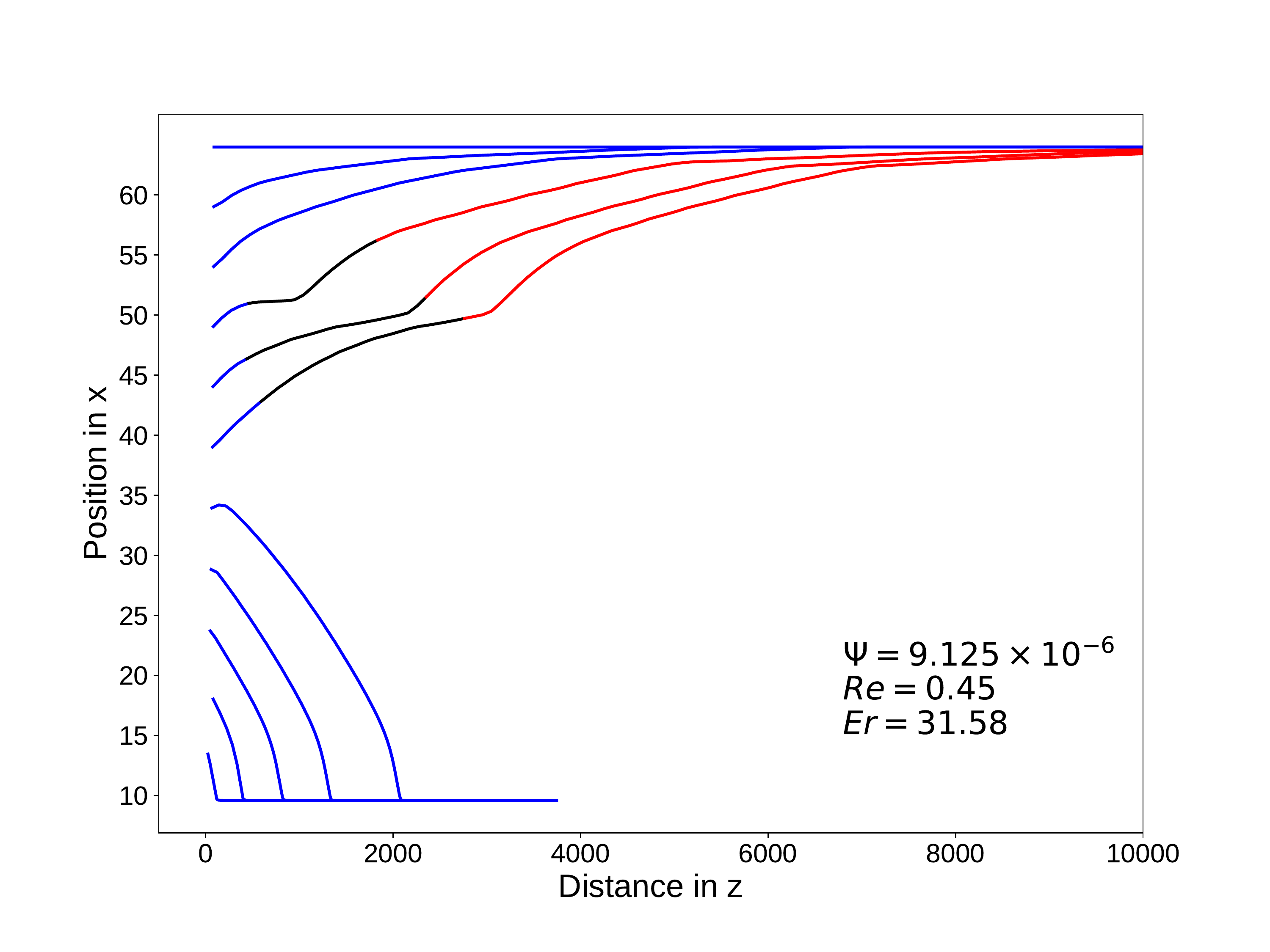}\\
\includegraphics[trim={2cm 1.2cm 2.1cm 2.6cm},clip, scale=0.23]{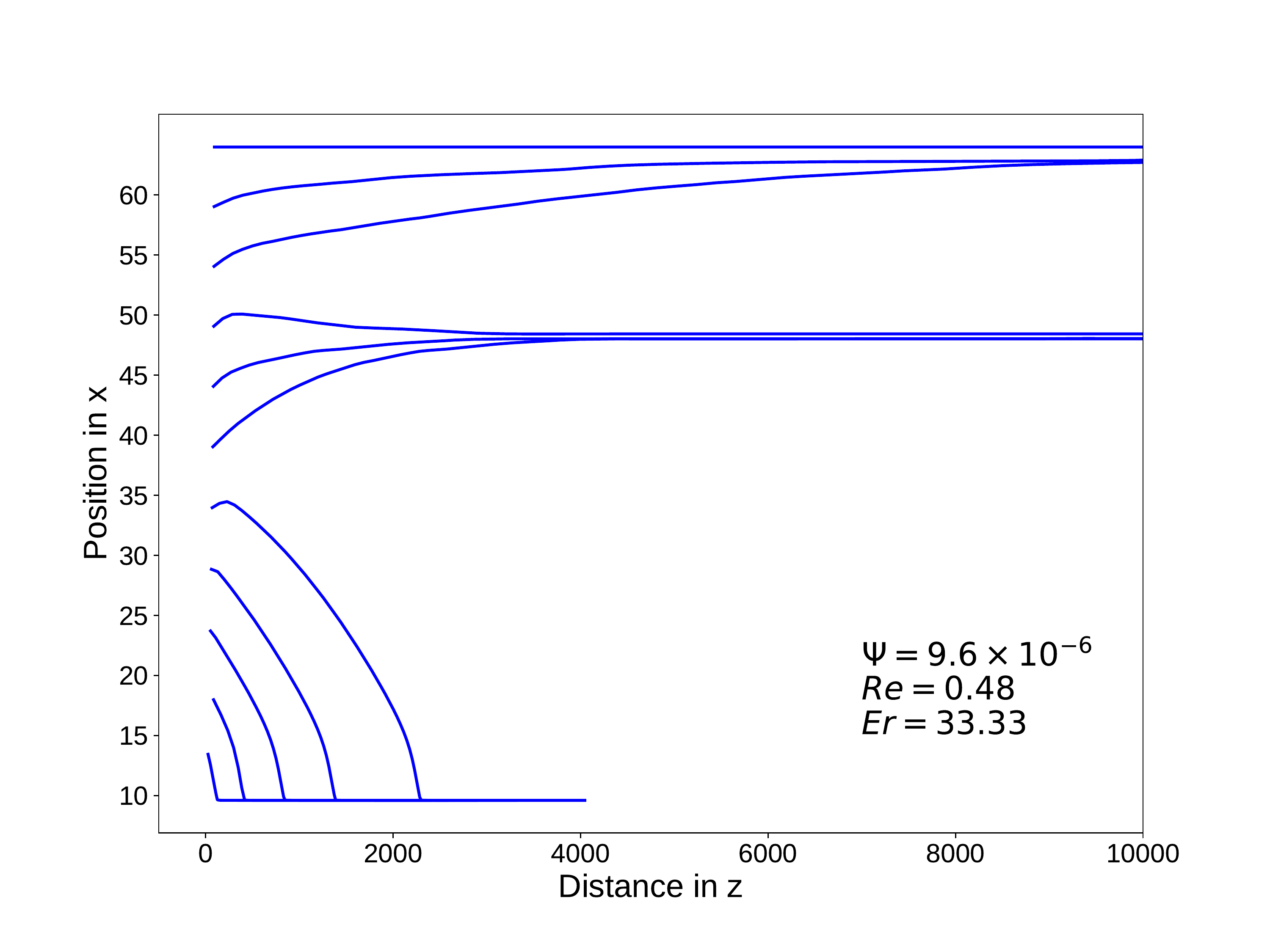}
\includegraphics[trim={2cm 1.2cm 2.1cm 2.6cm},clip, scale=0.23]{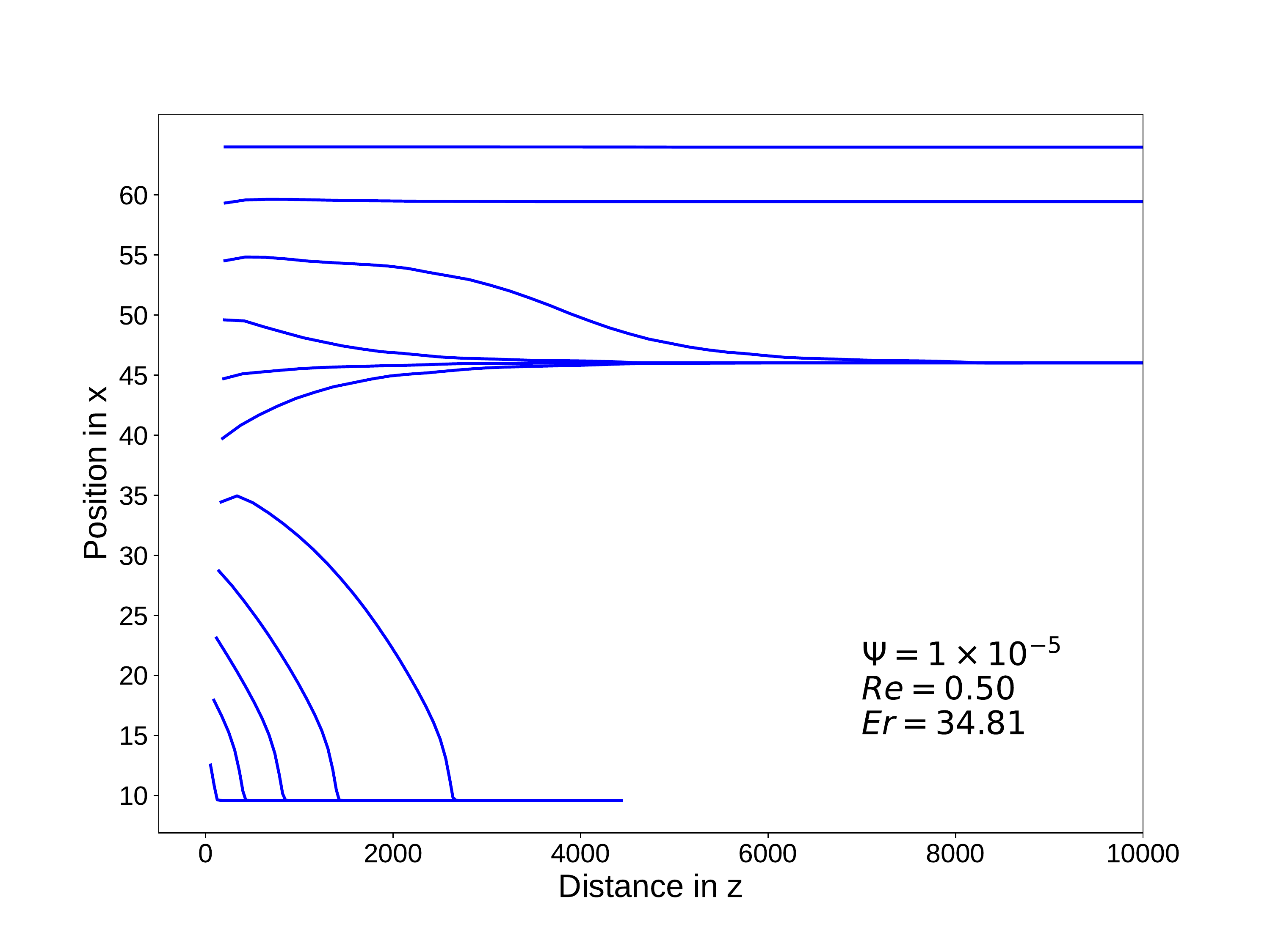}
\includegraphics[trim={2cm 1.2cm 2.1cm 2.6cm},clip, scale=0.23]{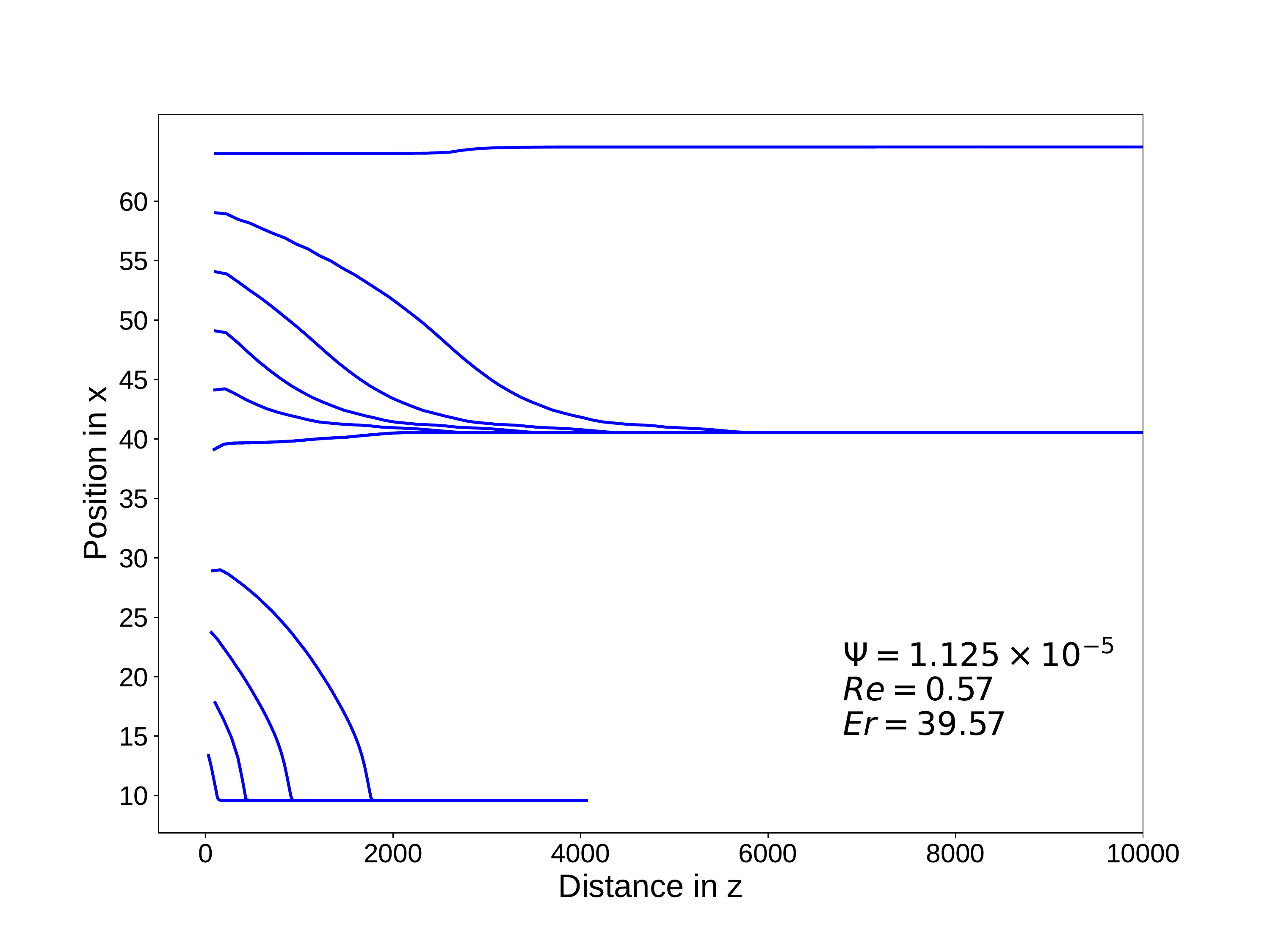}\\
\includegraphics[trim={2cm 1.2cm 2.1cm 2.6cm},clip, scale=0.23]{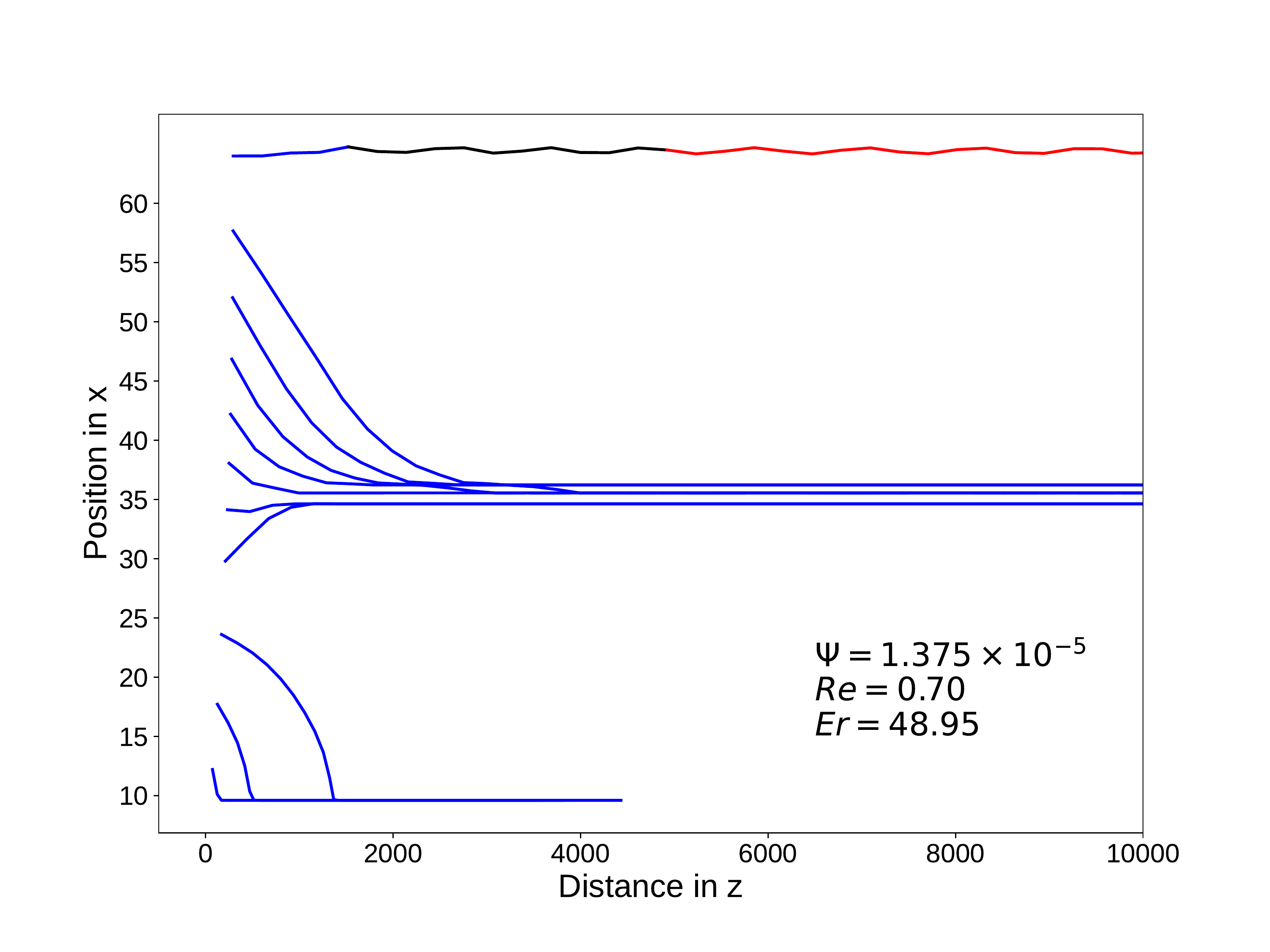}
\includegraphics[trim={2cm 1.2cm 2.1cm 2.6cm},clip, scale=0.23]{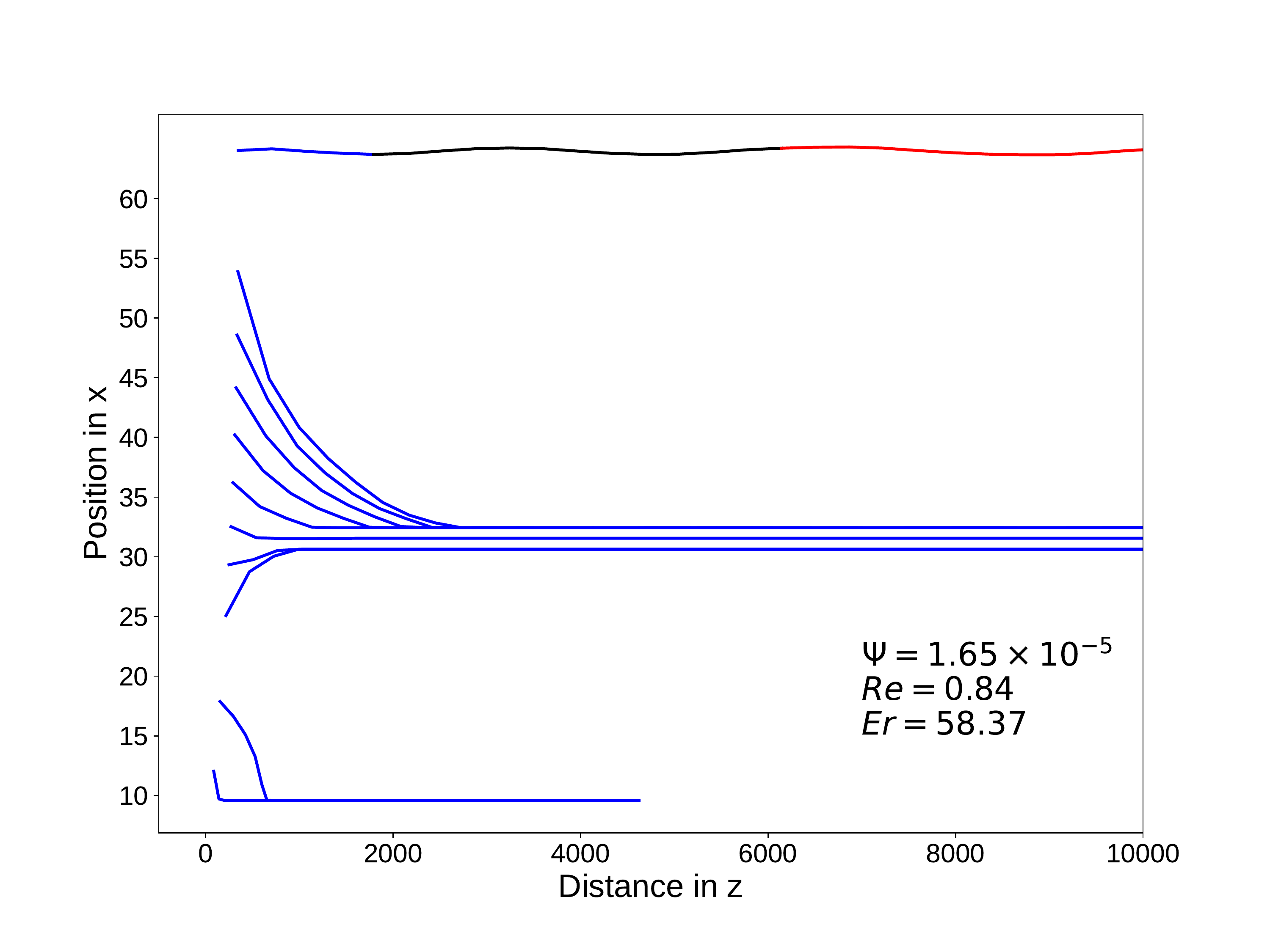}
\includegraphics[trim={2cm 1.2cm 2.1cm 2.6cm},clip, scale=0.23]{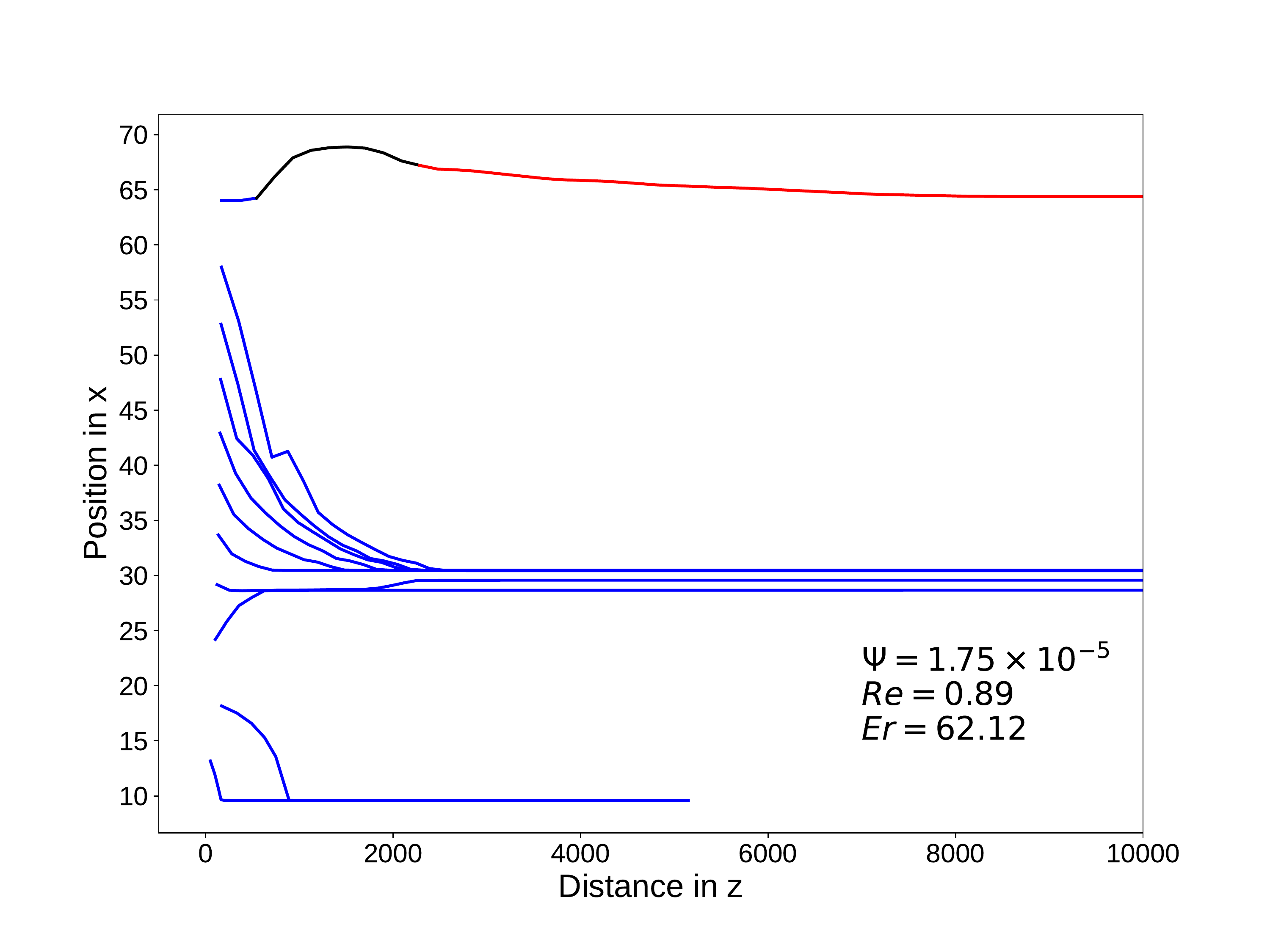}
\end{tabular}
\caption{Particle trajectories in a nematic liquid crystal host phase for particle size of $R=9.6$ and applied pressure gradients ranging from $\Psi=1.25 \times 10^{-6}$ to $1.75 \times 10^{-5}$. Blue lines indicate that the director structure is in a bend state, whereas red lines indicate that the director has transitioned to the splay state. Black lines mark a transition state between bend and splay. \rev{The particle Ericksen numbers $Er$ and particle Reynolds numbers $Re$ are given in each sub-plot. All quantities are given in LBU.
}}
\label{trajectories}
\end{figure*}

Fig.~\ref{trajectories} shows the complete set of trajectories for a particle with radius $R=9.6$ and pressure gradients ranging from $\Psi=1.25 \times 10^{-6}$ to $1.75 \times 10^{-5}$. The first row of Fig. \ref{trajectories} shows results for low pressure gradients $\Psi=1.25 \times 10^{-6},\,2.5 \times 10^{-6},\, 5 \times 10^{-6}$, indicating typical behaviour at low pressure gradients of  migration toward the wall or toward the weak attractor region that moves closer to the channel centre with increasing pressure gradient. In all cases the director remains in the bend state, as indicated by blue lines. 
The second row of Fig. \ref{trajectories} shows trajectories for intermediate pressure gradients $\Psi=7.5\times10^{-6},\,8.75\times10^{-6},\,9.125\times10^{-6}$, for which we can observe trajectories with overshooting and pull-back behaviour as well as a transition from the bend (blue lines) to the splay (red lines) state, through an intermediate state (black lines). For the lowest of these intermediate pressure gradients, $\Psi=7.5\times10^{-6}$, overshooting can be seen. 

Overshooting followed by pull-back to the centre is seen for $\Psi=8.75\times10^{-6}$, but the overshoot has disappeared at the higher pressure gradient of $\Psi=9.125\times10^{-6}$. In all cases of overshoot and pull-back there is a bend to splay transition.
The precursor of the emergent attractor is clearly seen in the trajectories for the  $9.125\times10^{-6}$ case, the right most image of the second row of Fig. \ref{trajectories}. For the highest intermediate pressure gradient, $\Psi=9.6\times10^{-6}$, the third attractor state emerges. The last two rows of Fig. \ref{trajectories} cover the higher range of pressure gradients, for which the emergent attractor state is possible, albeit for sufficiently high pressure gradients. It is clear that the position of the emergent attractor state moves towards the wall with increasing pressure gradient.
To further check the consistency of our findings, we also ran a number of additional simulations with $R=9.6$ at a lower shear and bulk viscosity $\eta=\zeta=1/6$, which led to the same equilibrium position and confirmed these results.

\begin{figure*}[htbp!]
\centering
\includegraphics[trim={0.75cm 1cm 1cm 1cm},clip, width=\linewidth]{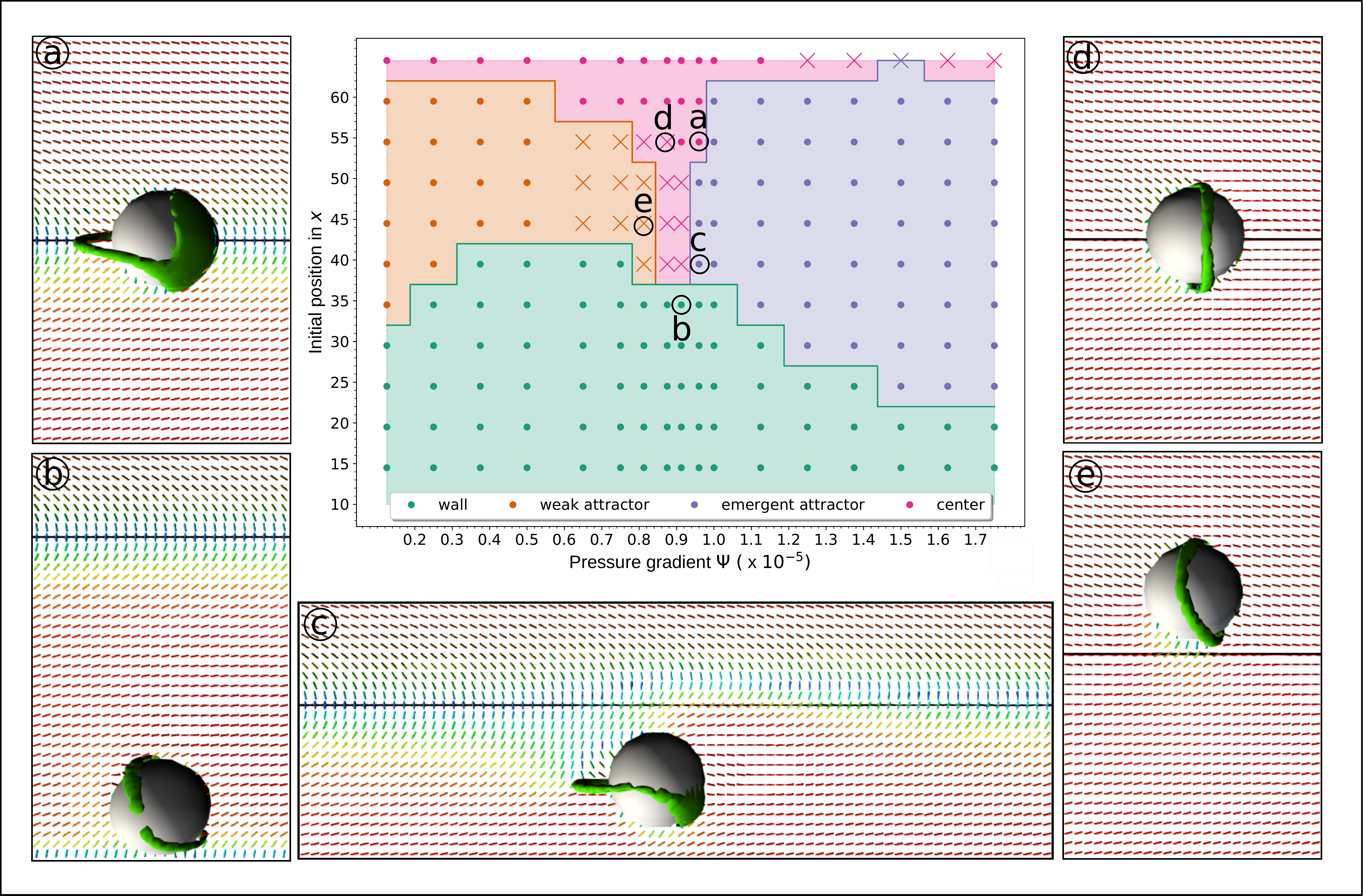}
\caption{Phase diagram of the preferential migration of the colloidal particle in a nematic host phase. Coloured regions show the equilibrium particle position (wall, weak attractor, emergent attractor, centre) as a function of the initial $x$-position of the particle and the applied pressure gradient $\Psi$. Equilibrium director bend states  are marked with dots, whereas equilibrium splay states are shown with crosses. Subplots (a)-(e) show the scalar order parameter (green isosurface showing the low order region) and director field (short coloured lines) around the particle for typical equilibrium states: (a) bend state, centre position; (b) bend state, wall position; (c) bend state, emergent attractor position;  (d) splay state, centre position; (e) splay state, weak attractor position. The centre of the channel is marked by the horizontal blue line. See also Movie.1, Movie.2, Movie.3, Movie.4 and Movie.5 in the electronic supplementary information for the dynamic evolution to the steady states (a), (b), (c), (d) and (e), respectively.}
\label{phase_diagram}
\end{figure*}

The main plot in Fig.~\ref{phase_diagram} summaries these results, depicting a phase diagram of the equilibrium position as a function of the starting position $x_s$ and the applied pressure gradient $\Psi$. For particles initially close to the walls, migration towards the walls occurs for all pressure gradients. Similarly, particles initially close to the centre of the channel migrate to the centre of the channel. Initial particle positions that are around one or two particle radii away from the centre give rise to a much more complicated behaviour: at low pressure gradients the particle migrates to an off-centre weak attractor position; for a narrow range of pressure gradients around $\Psi \simeq 9\times10^{-6}$ centre positions are again favoured; and above a critical pressure gradient $\Psi \simeq 9.5\times10^{-6}$ the phase diagram is increasingly dominated by the emergent attractor states. 

Also shown in Fig.~\ref{phase_diagram}, through the subplots (a)-(e), are the director and scalar order parameter structures within the channel. Previous theoretical and experimental work has shown that the director structure within channel flow of a nematic can exhibit either predominately bend or splay deformation \cite{Anderson2015, Copar2020} for low and high flow speeds, respectively. Both states are observed in our simulations, and both exhibit flow alignment at the appropriate Leslie angle, positive (negative) angles in the lower (upper) half of the channel where the shear gradient is positive (negative). As mentioned above, the states differ in their transition between positive and negative Leslie angle at the centre of the channel, with the bend state exhibiting director alignment along $x$ at the centre of the channel (for instance in Fig.~\ref{phase_diagram}(a)), and the splay state exhibiting director alignment along $z$ at the centre (Fig.~\ref{phase_diagram}(d)). For the pressure gradients we consider here, the bend state would normally be maintained. However, we observe a novel particle-induced mechanism for switching from the bend to splay state (see Fig.~\ref{bendsplaystate}) with the migration of the particle to the centre being the initiating event for a range of pressure gradients (see Fig.~\ref{phase_diagram}, indicated by crosses, and electronic supplementary information Movie.4 and Movie.5).

For wall equilibrium positions a bend state occurs with a tilted, but otherwise regular, Saturn ring defect around the particle (see Fig.\ref{phase_diagram}(b)). For centre equilibrium positions a bend state occurs for $\Psi < 1.2\times10^{-5}$ and a splay state for $\Psi > 1.2\times10^{-5}$.
For an initial particle position away from the centre, the bend state occurs for the weak attractor, centre and emergent attractor equilibrium positions, for both low and high pressure gradients. However, at intermediate pressure gradients $6\times10^{-6} \le \Psi \le 8.5\times10^{-6}$ we observe a transition to the splay state (Fig.\ref{phase_diagram} (d),(e)). We note that the transition from the bend to the splay state does not have a determining effect on whether the particles migrate to centre or off-centre positions as we see the same behaviour for initial positions close by and/or at lower pressure gradients. However, it affects the nature of the Saturn ring defect around the particle: for bend equilibrium states the Saturn ring defect is approximately horizontal  (Fig.~\ref{phase_diagram}(a) and (c)) and for splay equilibrium states the Saturn ring defect is approximately vertical  (Fig.~\ref{phase_diagram}(d) and (e)). For bend equilibrium states, the defect structure also develops a pronounced lip- or cap-shaped region of low liquid-crystalline order at the bow of the particle (Fig.~\ref{phase_diagram}(a),(c)). 

In comparison to the Segr\'e-Silberberg effect in isotropic fluids (Fig.~\ref{noLC}) we notice several fundamental differences. Firstly, in isotropic fluids, particles migrate to a single equilibrium position, whereas in nematic host phases migration to one of multiple equilibrium positions can occur, depending on particle starting positions and the applied pressure gradient. Secondly, while for the Segr\'e-Silberberg effect the location of the attractor state between the wall and centre depends only weakly on the flow velocity \cite{Segre1962,Schonberg1989} (in fact our simulations with an isotropic host phase, see Fig.~\ref{noLC}, show no appreciable change in equilibrium position over the entire range of applied pressure gradients), the position of the emergent attractor state in the nematic system depends much more sensitively on the imposed pressure gradient, and therefore on the Reynolds number. For instance, the Reynolds numbers for which there are emergent attractor states in Fig.~\ref{attractor} are  $Re\simeq0.48$ and $Re\simeq0.57$ and, even with this small increase in $Re$, the attractor position moves by almost one particle radius. Finally, the preferential migration in a nematic host phase happens more than an order of magnitude faster than in isotropic fluids. For instance, for a pressure gradient $\Psi=1.125\times10^{-5}$ (Fig.~\ref{attractor}, green lines), where the position of the emergent attractor in the nematic host phase and the Segr\'e-Silberberg equilibrium position in the isotropic host phase almost coincide, particles in a nematic host reach their equilibrium positions by the time they have travelled around $z \simeq 5\times10^3$ along the channel, and in an isotropic host phase (see Fig.~\ref{noLC}) take, depending on the start positions $x_s$, at least an order of magnitude longer (in time or distance along the channel).  

\rev{With regard to Ericksen numbers $Er$, our results indicate that the observed pattern of preferrential migration occurs in a regime where viscous forces are larger than elastic forces, and where consequently the director field is strongly affected by the flow field. At the lower end of this regime, for $3.43\le Er \le 16.55$, the weak attractor state exists, as shown in the top row of the Fig.~\ref{trajectories}. The emergent attractor state occurs for $Er\simeq 30$ and above. With increasing $Er$, viscous forces begin to dominate over elastic forces. This means the flow behaviour at higher Ericksen numbers is more akin to that of an isotropic fluid, which shows the classical Segr\'e-Silberberg effect. However, the fact that a strongly flow-aligned liquid crystal forms the host phase leads always to certain qualitative differences. For instance, the trajectories shown in the two bottom rows of Fig.~\ref{trajectories} feature the emergent attractor state at $33.33\le Er\le62.12$, do show some similarities to the classic Segr\'e-Silberberg effect. However, the movement of the emergent attractor region towards the walls with increasing Ericksen number, the attraction to the walls, or the existence of stable trajectories at the channel centre are all features that arise due to the anisotropic nature of the host phase and the interaction of flow-aligned director field with the defect structure around the particle, as is visible in the snapshots shown in Fig.~\ref{phase_diagram}(a)-(e).}

\begin{figure}[htbp!]
\centering
\includegraphics[trim={1.5cm 1cm 2.8cm 2.5cm},clip,width=\linewidth] {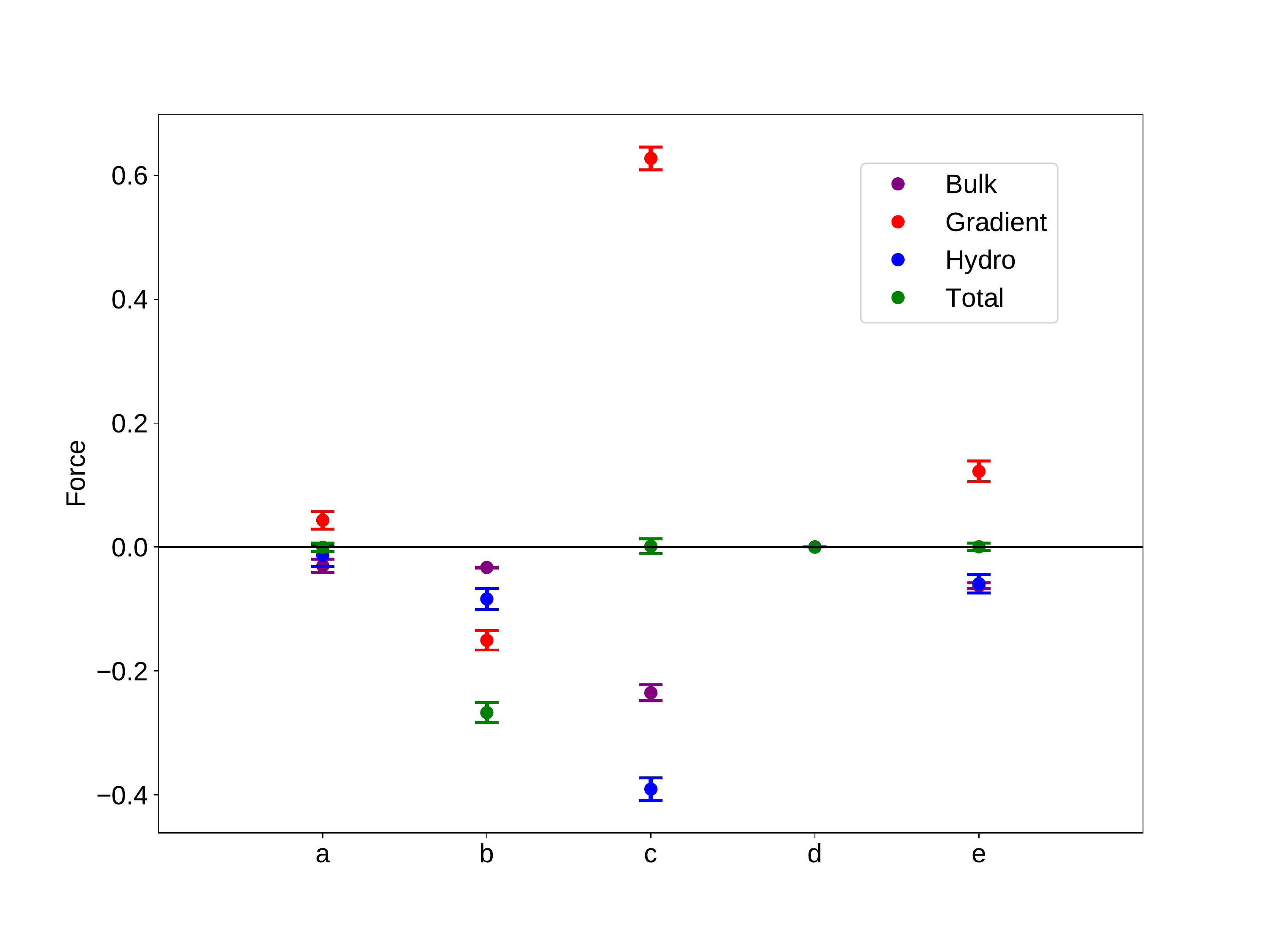}
\caption[]{Total force and the contributions to the total force on the particle in the steady state for particles that have equilibrated to lie at (a) the centre with asymmetric defect, (b) at the wall, (c) at the emergent attractor, (d) at the centre with symmetric defect, and (e) at the weak off-centre attractor, for particle radius $R=9.6$. The error bars indicate one standard deviation of the force data during a run over $10^4$ algorithmic steps. All quantities are given in LBU.}
\label{force_comp}
\end{figure}

In order to investigate the presence of this emergent attractor, we consider the total force and the contributions to the total force on the particle in the steady state for particles that have equilibrated (see Fig. \ref{force_comp}). An analysis of the individual force contributions shows that, for a centre equilibrium position as in Fig.~\ref{phase_diagram}(d) all three force components vanish, as expected from symmetry, and for a wall equilibrium position as in Fig.~\ref{phase_diagram}(b) all three force components are negative, forcing the particle to remain at the wall. 
For particle migration to an emergent attractor state as in Fig.~\ref{phase_diagram}(c), our simulation show that the particle feels a force towards the channel centre from the gradient, i.e.~elastic, terms and forces towards the wall from both the bulk term and the hydrodynamic force component.
This is also the case for particle migration to the centre with an asymmetric defect as in Fig.~\ref{phase_diagram}(a) and for particle migration to the weak attractor as in Fig.~\ref{phase_diagram}(e), although the individual force contributions are significantly smaller. Therefore, a delicate force balance exists between these relatively large contributions leading to a zero total force at equilibrium. 

\begin{figure}[htbp!]
\centering
\includegraphics[trim={2cm 1.2cm 0.5cm 2.2cm},clip, width=\linewidth] {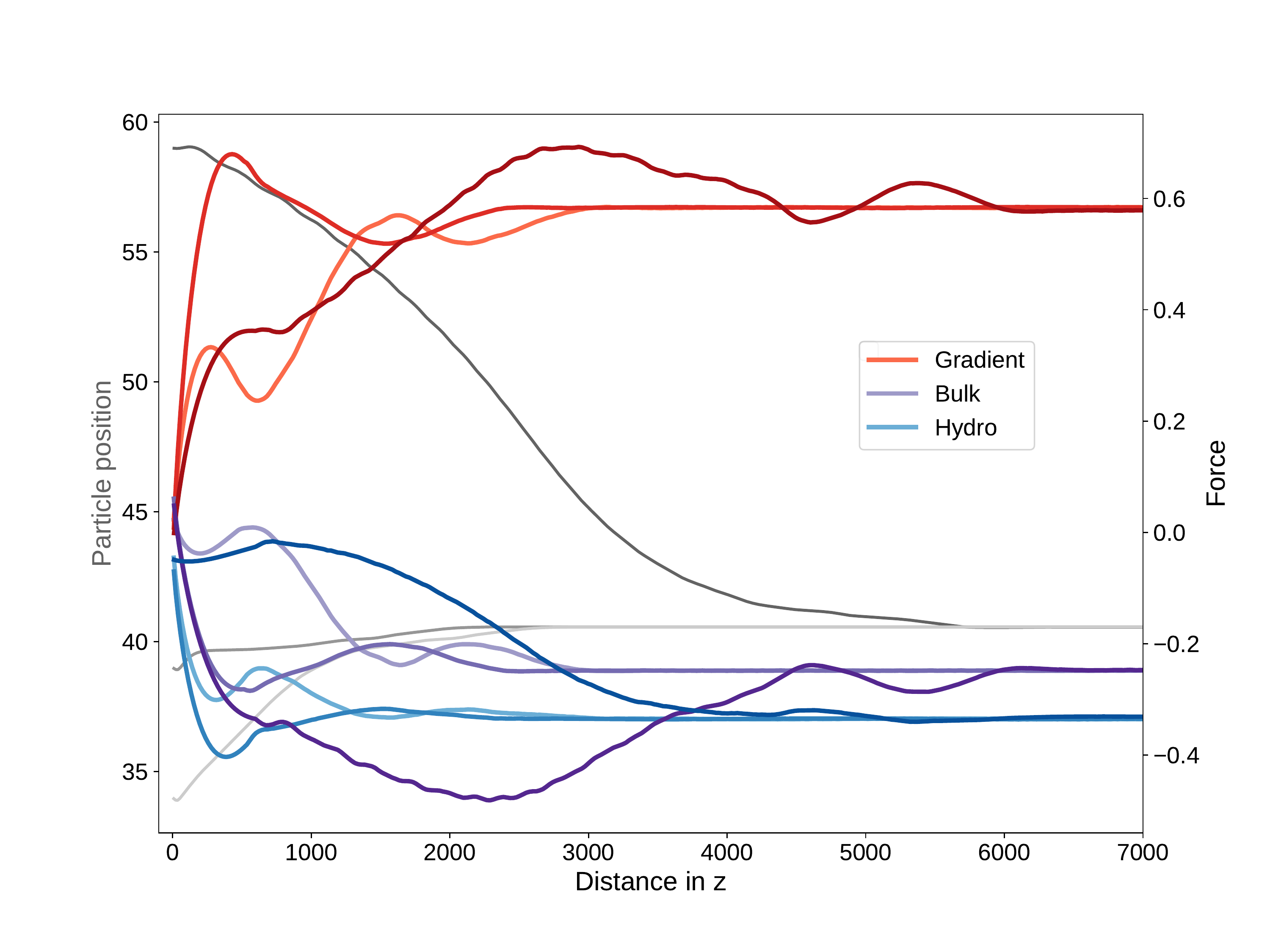}
\caption[]{Time evolution of force contributions in the emergent attractor state with pressure gradient $\Psi=1.125 \times 10^{-5}$. Trajectories for three starting positions are shown, for $x=34.5$ (light grey), $x=39.5$ (medium grey) and $x=59.5$ (dark grey). Also shown are the contributions to the total force on the particle which arise from gradient (red), bulk (purple) and hydrodynamic (blue) terms  with the hue (light, medium, dark) corresponding to the equivalent starting position (light grey, medium grey, dark grey). All quantities are given in LBU.}
\label{force_vs_time}
\end{figure}

Fig.~\ref{force_vs_time} shows the time evolution of the different contributions to the force on the particle during its approach to an emergent attractor state, for pressure gradient $\Psi=1.125\times10^{-5}$ and particle radius $R=9.6$. For these values of $\Psi$ and $R$ the attractor state is located, approximately, at  the Segr\'e-Silberberg  equilibrium $x$-position (namely that seen in  Fig. \ref{noLC}). 
Trajectories for three starting positions are shown, for $x=34.5$ (light grey), $x=39.5$ (medium grey) and $x=59.5$ (dark grey), which cover a range of migration patterns to the attractor from below and above. Also shown are the contributions to the total force on the particle which arise from gradient (red), bulk (purple) and hydrodynamic (blue) terms  with the hue (light, medium, dark) corresponding to the equivalent starting position (light grey, medium grey, dark grey).

From Fig.~\ref{force_vs_time} we observe initial transient behaviour before the particle reaches the same equilibrium position for all three starting positions. For all starting positions and for all time, the gradient contribution is positive, so that the elastic forces always act to  move the particle towards the channel centre. For both bulk and hydrodynamic force contributions are negative for all starting positions and (almost all) time, so that they act to move the particle towards the wall. The exception is that for starting position $x=34.5$ the bulk force contribution is positive for a very short initial period, i.e.~for very small $z$-displacements of the particle from the initial position. 
Interestingly, all three force contributions are almost balanced, with a total force of zero, for all time. Variations in the total force are indiscernible on the same force scale used in Fig.~\ref{force_vs_time}.  

This situation is in contrast to the Segr\'e-Silberberg effect, in which the \rev{inertial} component of the hydrodynamic force acts to move the particle across the shear gradient towards the wall, while the increased pressure caused by the particle moving towards the wall leads to a force acting to move the particle towards the centre of the channel. At the Segr\'e-Silberberg equilibrium position, the total hydrodynamic force vanishes. In a nematic host material, the gradient terms act with the \rev{inertial} hydrodynamic forces to move the particle towards the wall, allowing much faster migration of the particle and the appearance of an attractor state at a much smaller Reynolds number. 

\begin{figure*}
\centering
\begin{tabular}{ccc}
\includegraphics[trim={2cm 1.2cm 2.1cm 2.6cm},clip, scale=0.23]{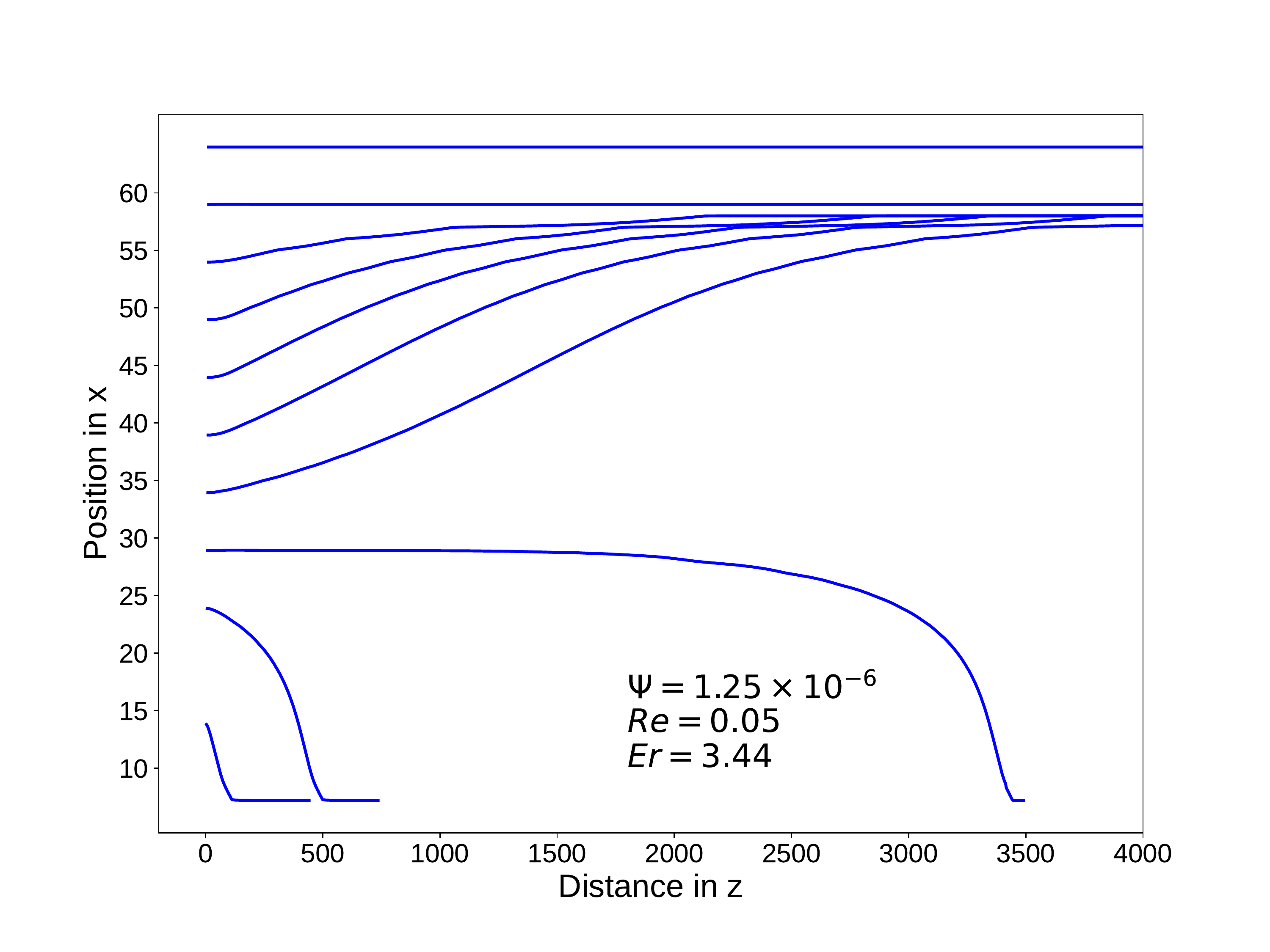}
\includegraphics[trim={2cm 1.2cm 2.1cm 2.6cm},clip, scale=0.23]{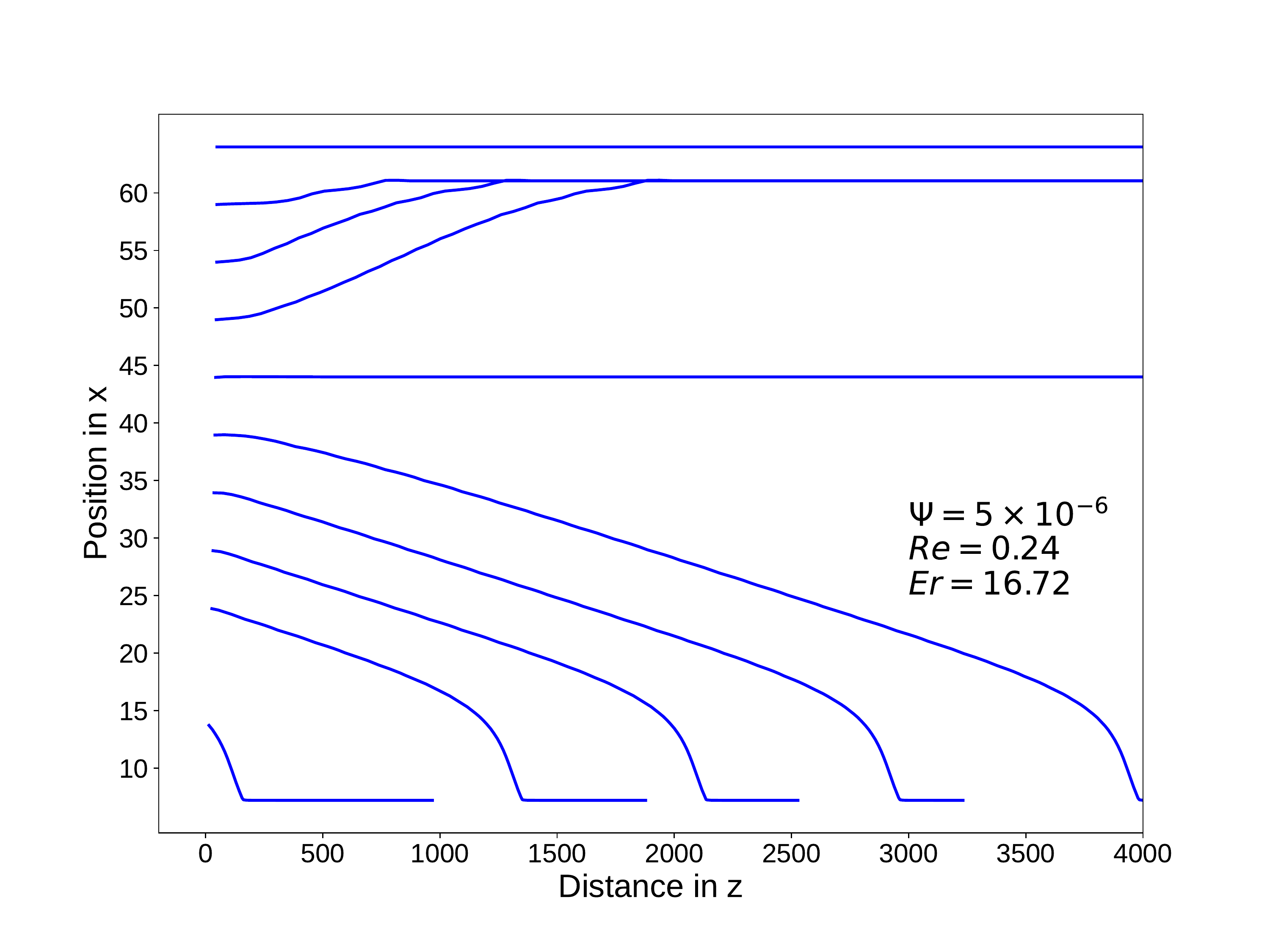}
\includegraphics[trim={2cm 1.2cm 2.1cm 2.6cm},clip, scale=0.23]{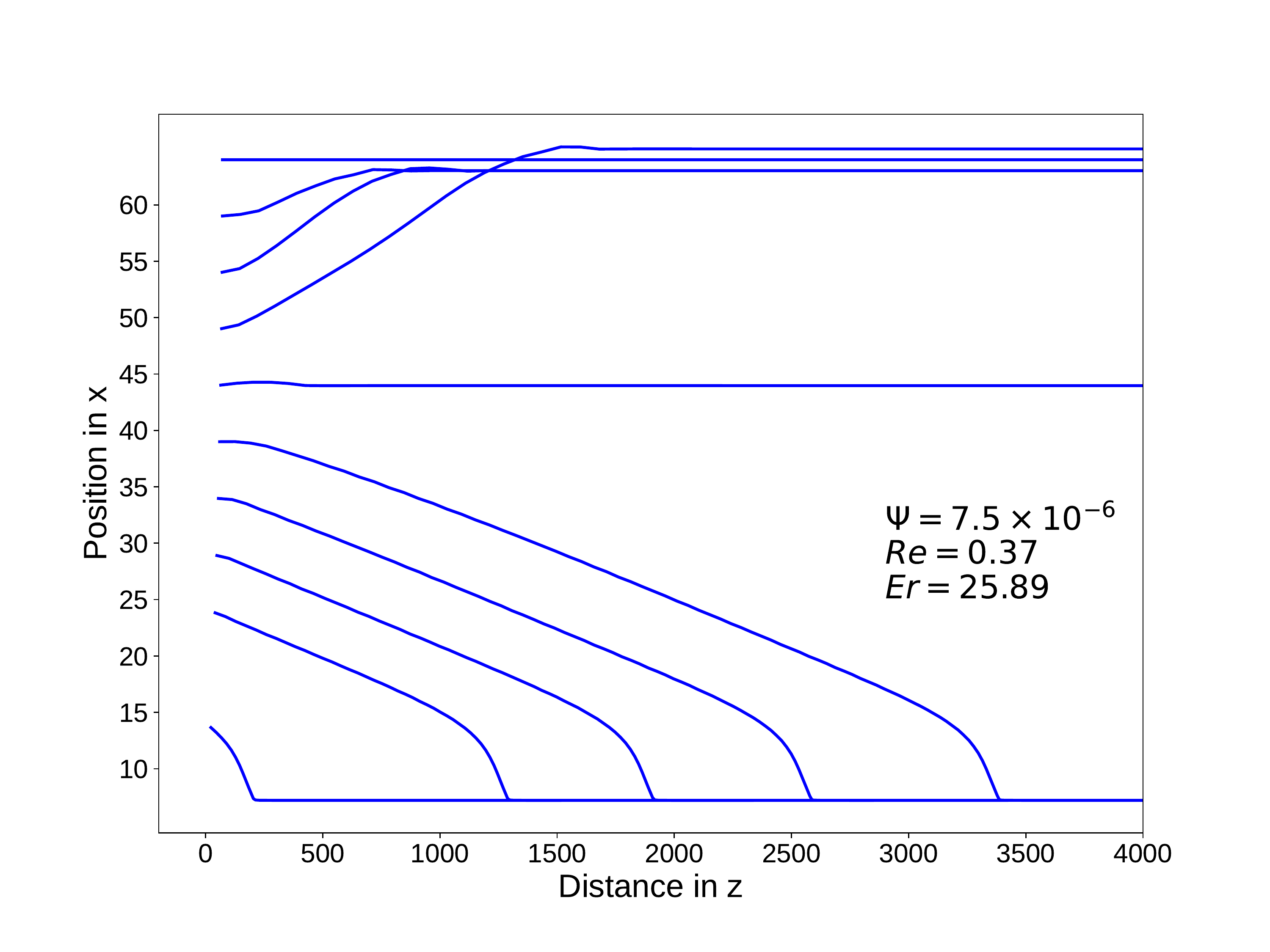}\\
\includegraphics[trim={2cm 1.2cm 2.1cm 2.6cm},clip, scale=0.23]{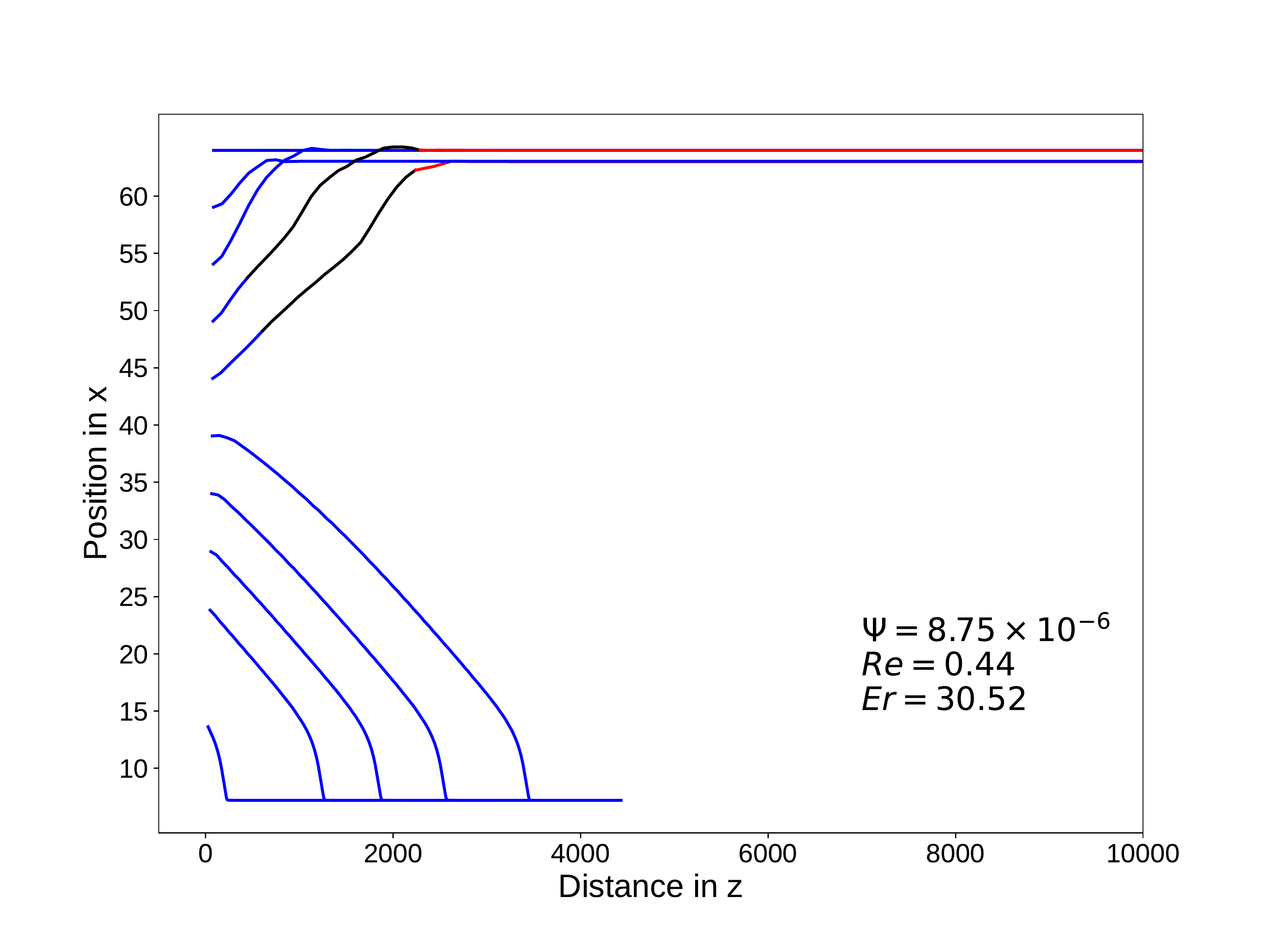}
\includegraphics[trim={2cm 1.2cm 2.1cm 2.6cm},clip, scale=0.23]{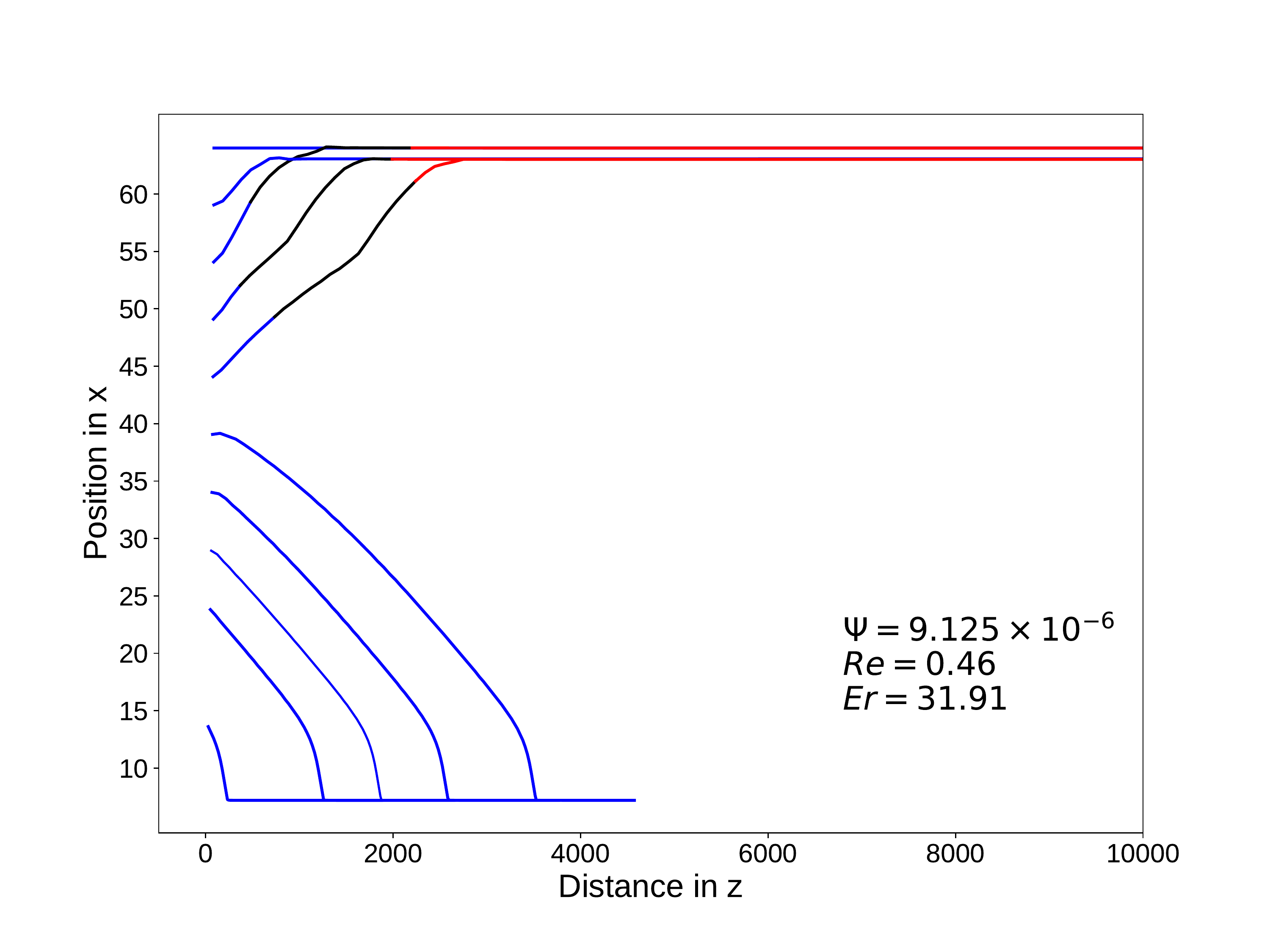}
\includegraphics[trim={2cm 1.2cm 2.1cm 2.6cm},clip, scale=0.23]{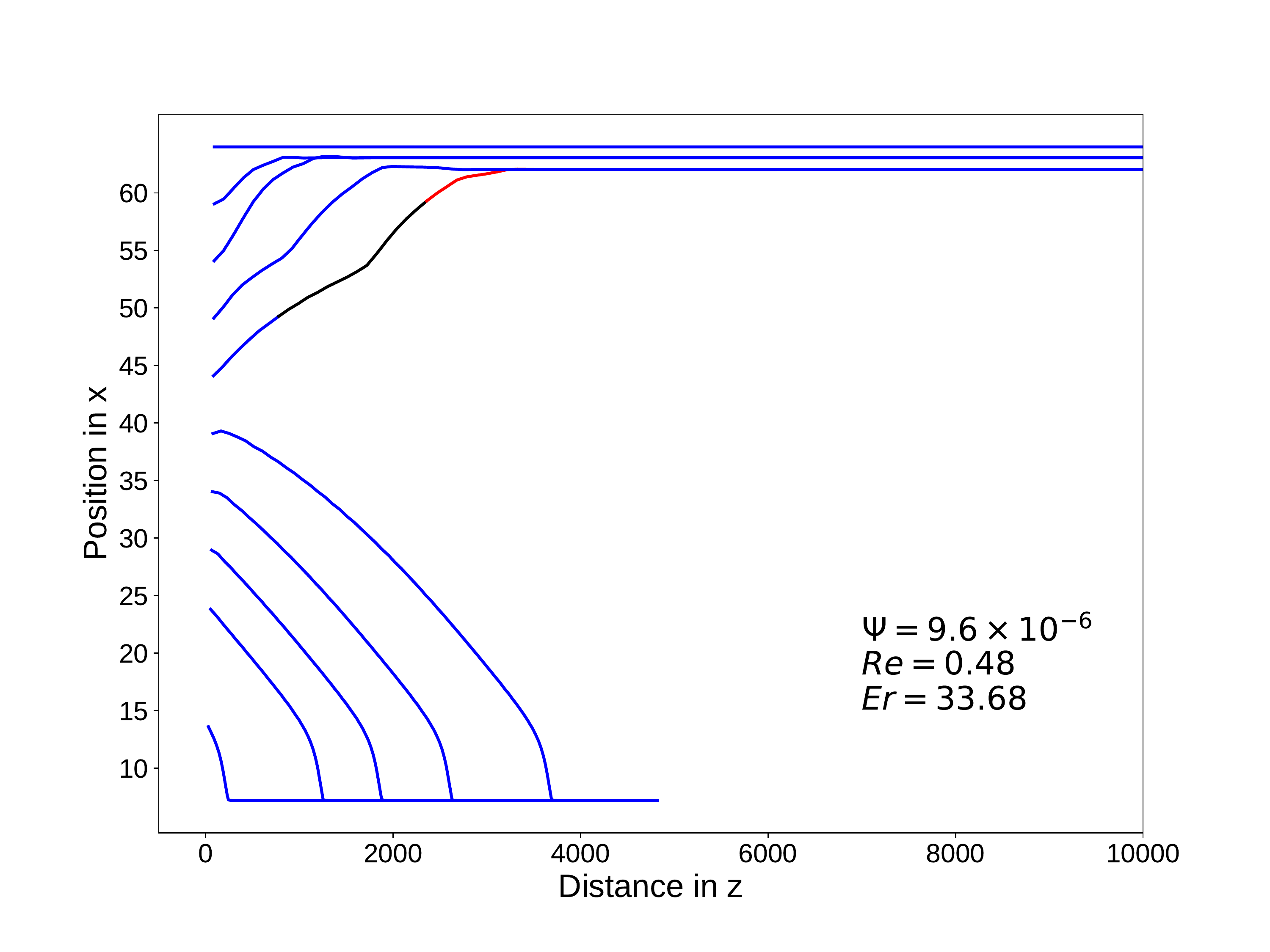}\\ 
\includegraphics[trim={2cm 1.2cm 2.1cm 2.6cm},clip, scale=0.23]{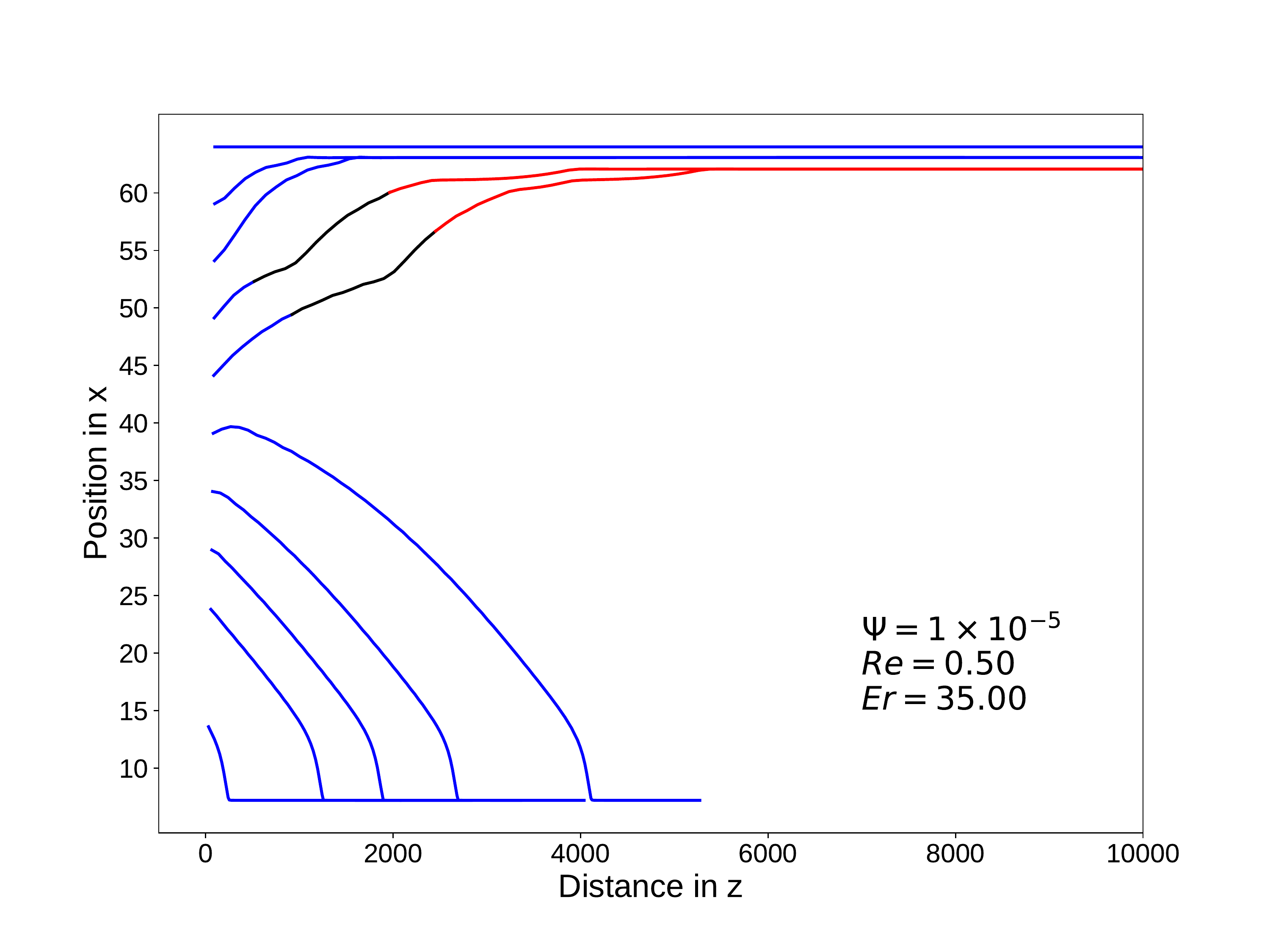}
\includegraphics[trim={2cm 1.2cm 2.1cm 2.6cm},clip, scale=0.23]{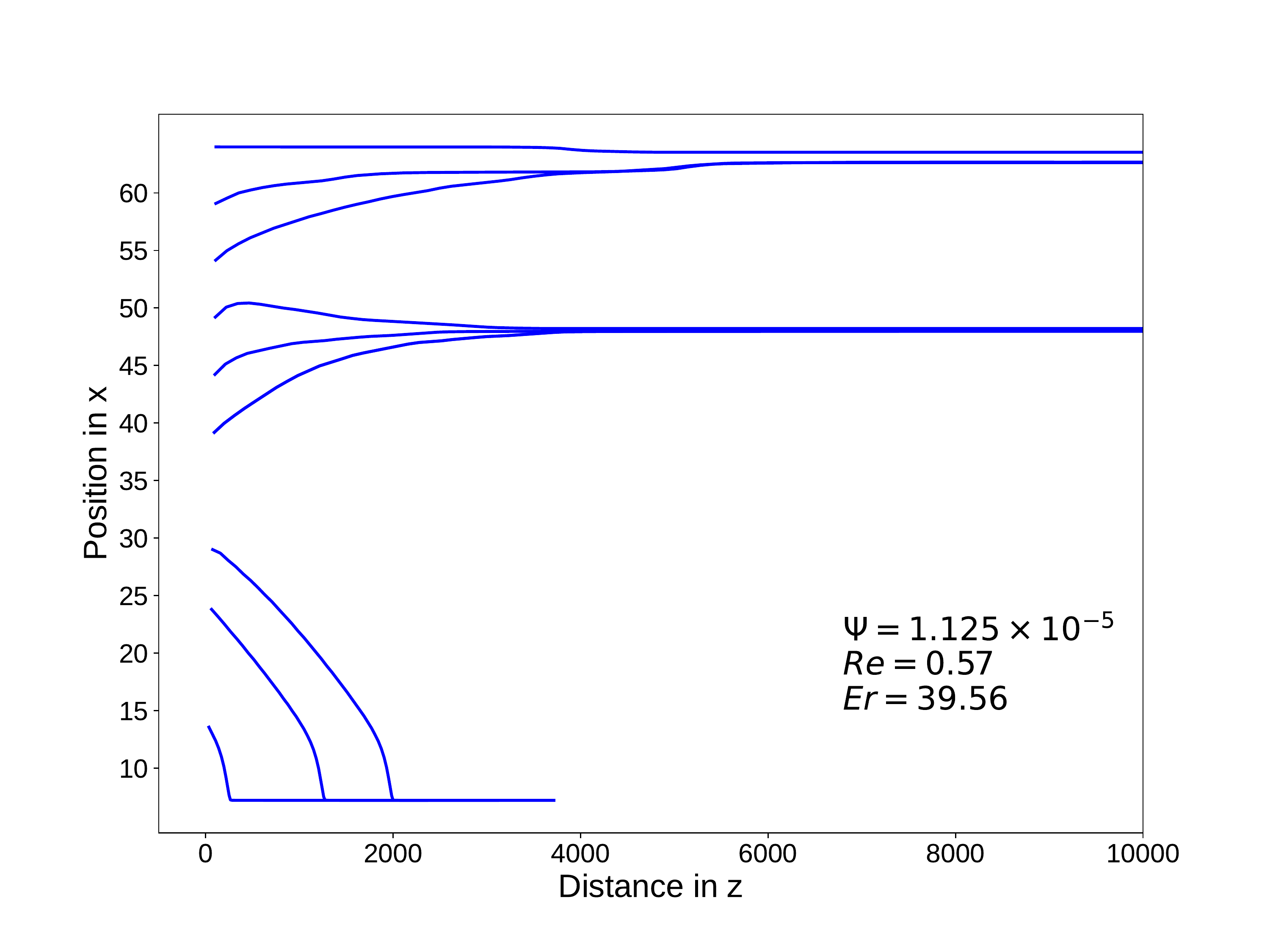}
\includegraphics[trim={2cm 1.2cm 2.1cm 2.6cm},clip, scale=0.23]{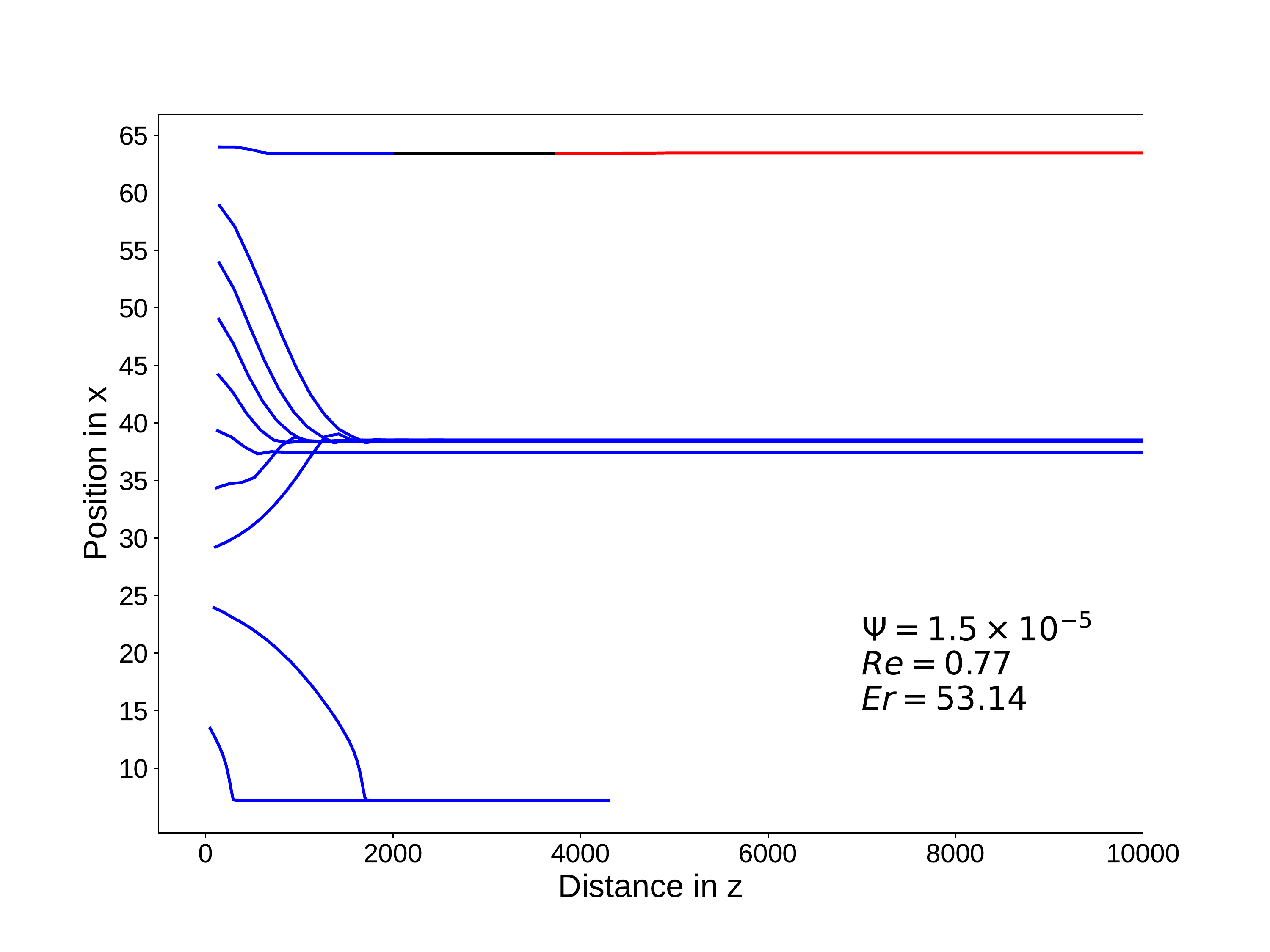} 
\end{tabular}
\caption{Particle trajectories in a nematic liquid crystal host phase for  particle size $R=7.2$ and various applied pressure gradients from. Blue lines indicate that the director structure is in a bend state, red lines indicate the splay state and black sections denote the transition. \rev{The particle Ericksen numbers $Er$ and particle Reynolds numbers $Re$ are given in each sub-plot. All quantities are given in LBU.
}}
\label{trajectories_7_2}
\end{figure*}

The presence of a particle also leads to a distorted director structure, rather than a uniform director at the Leslie angle and so increased pressure gradients (equivalent to an increased Reynolds number) can align the director around the particle more closely to the Leslie angle, thus adapting the elastic force on the particle, and therefore control of the attractor equilibrium $x$-position through changes in pressure gradient is possible.

Additional results for smaller particles with $R=7.2$ show a similar pattern of preferential migration, and are depicted Fig.~\ref{trajectories_7_2} as trajectories. There are quantitative differences, for instance overshooting off-centre trajectories or trajectories with overshooting and pull-back to the channel centre are much less pronounced compared to larger particles with $R=9.6$. This can be attributed to smaller inertia and reduced anchoring forces, which scale down with the surface area of the particle. For smaller particles the attractor state emerges at slightly larger pressure gradient around $\Psi=1.125\times10^{-5}$, but shows the same characteristics, in particular the fast migration of the particle to the equilibrium regions, as well as the movement of the emergent attractor position towards the walls and its prominence in the phase diagram with increasing pressure gradient.
These results suggest that this new effect depends to a certain extent also on particle size, 
\rev{and therefore on inertia. For higher inertia we expect the emergent attractor region in the phase diagram Fig.~\ref{phase_diagram} and states as in Fig.~\ref{phase_diagram}(c) to extend towards lower pressure gradients. The middle regions with states as in Fig.~\ref{phase_diagram}(d) and (e) are also likely to appear at lower pressure gradients  and to grow in size. This would occur at the cost of the region with wall attraction and states as in Fig.~\ref{phase_diagram}(b), which is likely to have a smaller extent.}

\section{Conclusions}

In summary, we observe multiple equilibrium particle positions and a new pressure-controllable particle attractor state for a colloidal particle with nematic liquid crystalline host phase. At low pressure gradients particles migrate either to the channel centre or the walls but at higher pressure gradients a third attractor state emerges spontaneously, whose position in the channel depends sensitively on the pressure gradient. These results are in  striking contrast to the classical  Segr\'e-Silberberg effect in isotropic fluids, where the equilibrium position is also reached more slowly. The discovery of these new and controllable attractor positions opens up interesting routes for tailored particle separation. 
\rev{While our results were obtained in pressure-driven flow, we expect them to hold as well in flux-driven flow as long as there is no significant drag between particle and fluid.} 
However, it is likely that, as well as particle size, the confinement ratio and  anchoring type and strength offer additional mechanisms to control the particle migration. These points will be addressed in future studies.

\section*{Conflicts of Interest}

There are no conflicts to declare.

\section*{Acknowledgements}

We acknowledge support from EPSRC under Grant No. EP/N019180/2 (O.H.) and EP/R513349/1 (M.L.). M.L. acknowledges funding from the Mac Robertson Postgraduate Travel Scholarship.
For the purpose of complying with UKRI's open access policy, the author has applied a Creative Commons Attribution (CC BY) license to any Author Accepted Manuscript version arising from this submission.
This work used the ARCHIE-WeSt High Performance Computer (www.archie-west.ac.uk) based at the University of Strathclyde and the Cirrus UK National Tier-2 HPC Service (http://www.cirrus.ac.uk).

\section*{Data Availability}

\rev{Data underpinning this publication are openly available from the University of Strathclyde KnowledgeBase research information portal.}

\bibliography{references}
\bibliographystyle{rsc}

\end{document}